%% file: intraover.tex
\documentclass[12pt]{article}

\input{common_defs}
\begin{document}

\title{\paperTitle
  \thanks{We thank seminar participants at Northwestern University, UT Austin,
    University of Florida Conference on Machine Learning in Finance, Society of Quantitative
    Analyst Data Science in Finance Conference, Cubist Systematic Strategies, Wolfe Research NLP
    Conference, Stony Brook Quantitative Finance Conference, Peking University, INFORMS Annual
    Meeting, Institute for Mathematical and Statistical Innovation, CIBC, and discussants Farid
    Aitsahalia and Yin Luo.}}

\authorLine

\maketitle

\begin{abstract}
  Over the past 30 years, nearly all the gains in the U.S. stock market have been earned overnight,
  while average intraday returns have been negative or flat. We find that a large part of this
  effect can be explained through features of intraday and overnight news. Our analysis uses a
  collection of 2.4 million news articles. We apply a novel technique for supervised topic analysis
  that selects news topics based on their ability to explain contemporaneous market returns.  We
  find that time variation in the prevalence of news topics and differences in the responses to news
  topics both contribute to the difference in intraday and overnight returns.  In out-of-sample
  tests, our approach forecasts which stocks will do particularly well overnight and particularly
  poorly intraday. Our approach also helps explain patterns of continuation and reversal in intraday
  and overnight returns. We contrast the effect of news with other mechanisms proposed in the
  literature to explain overnight returns.
\end{abstract}


\thispagestyle{empty}
\newpage
\setcounter{page}{1}

\section{Introduction}\label{s:intro}

Several research studies and accounts in the popular press have reported that overnight gains in the
U.S. stock market outstrip intraday returns. Indeed, over at least the past 30 years, nearly all
gains in the market have been earned overnight. Past research has also documented patterns of
continuation and reversal specific to intraday and overnight returns. Prior work has highlighted
trading by retail investors, institutional investors, and financial intermediaries related to these
phenomena.

\begin{figure}
\centering
\includegraphics[width=1\linewidth]{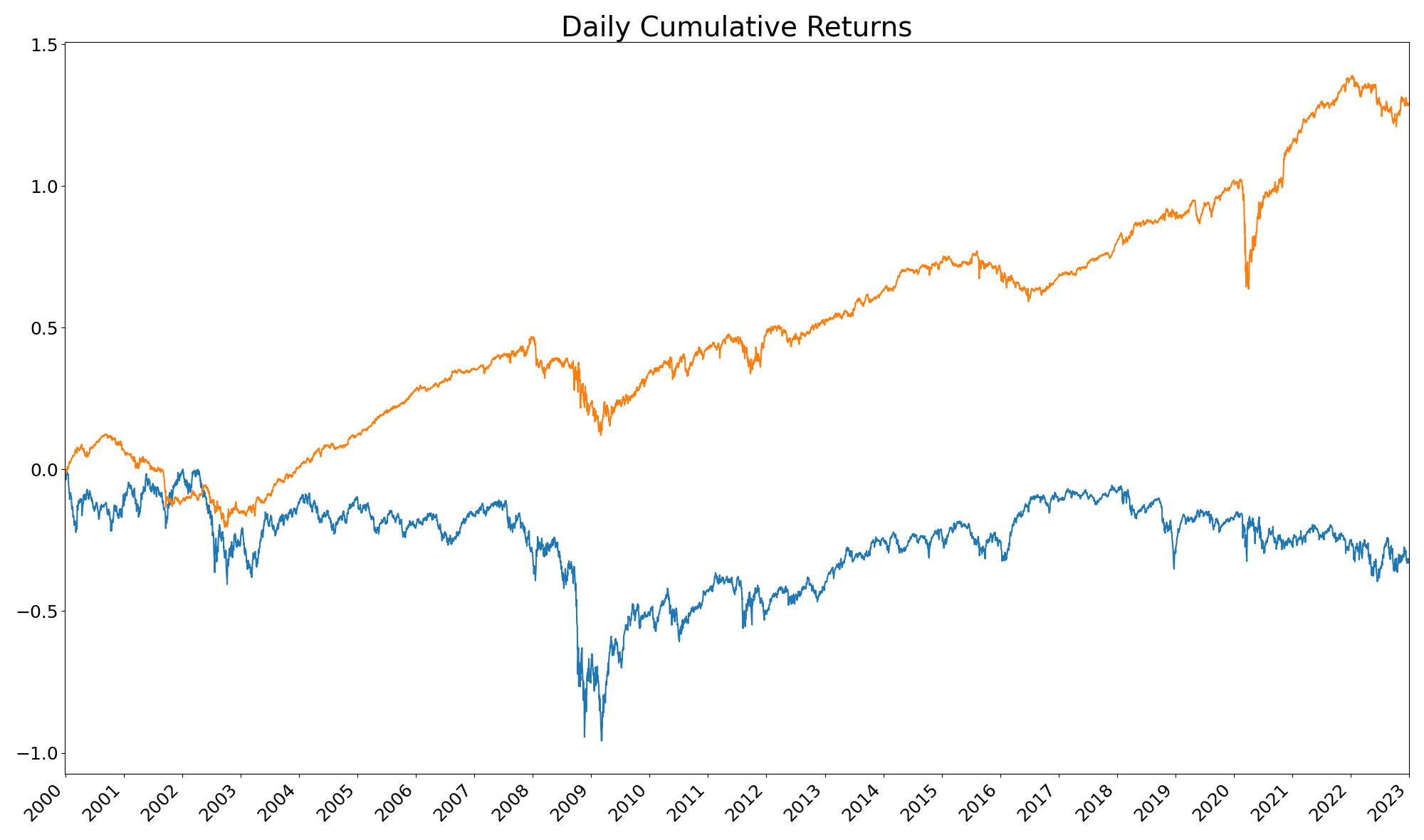}
\caption{Cumulative intraday and overnight daily log returns of S\&P 500 stocks, equally weighted, 2000--2022.}
\label{f:cum_intraover}
\end{figure}

To illustrate, Figure~\ref{f:cum_intraover} shows that cumulative overnight returns of S\&P 500
stocks have dramatically exceeded cumulative intraday returns since 2000.  The difference
  averages to 2.75 basis points per day, or roughly 7.2\% per year. Table \ref{t:io-mom-rev} shows
correlations between intraday and overnight returns over consecutive years. We will refer to the
patterns in this table (to which we return shortly) as \emph{over-intra correlations}.

We find an important role for the flow of news in explaining patterns of intraday and overnight
returns. We start with a large dataset of the full text of news articles from 1996 to 2022. We use
topic analysis to characterize the content of the news text, and we measure the exposure of each
S\&P 500 company to each topic in each period (intraday and overnight) on each day. We find
differences in the prevalance of news topics in the intraday and overnight periods; importantly, we
also find differences in the market response to a company's exposure to news topics between the two
periods.

Using this framework, we forecast a stock's future returns, intraday and overnight, based on past
exposure to news. In our main specification, we forecast returns one year ahead based on four years
of past news. We find that our top overnight picks substantially outperform the overall overnight
market, and our bottom intraday picks substantially underperform the overall intraday
market. Moreover, the remaining intraday-overnight difference in performance becomes statistically
insignifcant once we remove our top and bottom picks. (This holds whether we measure performance
through average returns or abnormal returns relative to standard factors. It also holds whether
returns are measured over the same years used to estimate the topic model or over later years.)
Thus, our news analysis helps explain the difference in returns across the two parts of the day.

With annual returns (calculated separately for intraday and overnight periods), we find return
continuation intraday to intraday and overnight to overnight, negative contemporaneous correlation
between intraday and overnight returns, and, as a consequence, patterns of reversal from one period
(intraday or overnight) to the other; see Table \ref{t:io-mom-rev}.  These patterns are partly
explained by differences in the prevalence and response to news topics because companies' topic
exposures are persistent and the intraday and overnight market responses to news topics are
generally negatively correlated. We find that including our news-based return forecasts diminishes
the impact of lagged returns in predicting future returns: news flow partly explains the over-intra
correlations in Table \ref{t:io-mom-rev}.

Because of the extreme turnover required to trade the over-intra effect, our findings fall short of
being a viable trading strategy. However, they might still be useful for institutional investors
concerned with optimal timing of market entry and exit, as well as for market makers who can
preposition their inventories to account for the over-intra effect.

\begin{table}
  \centering
  {\bf Annual intraday and overnight return momentum and cross-reversals} \\[5pt]
  \begin{tabular}{lrrr}
    \toprule
    {} & $r^{intra}_{t+1}$ & $r^{over}_{t+1}$ & $r_{t+1}$ \\
    \midrule
    $r^{intra}_t$ &           0.19*** &         -0.27*** &           \\
    $r^{over}_t$  &          -0.26*** &          0.29*** &           \\
    $r_t$         &                   &                  &     -0.04 \\
    \bottomrule
  \end{tabular}
  \caption{Correlations between intraday and overnight returns in years $t$ and $t+1$, 2000--2021,
    showing continuation and reversal effects, using stacked firm-year observations. Close-to-close
    returns $r_t$ show neither. 
    $^{***}$p$<$0.01, using standard errors clustered by year and firm.}
  \label{t:io-mom-rev}
\end{table}

\subsection{Literature review}

Using S\&P 500 data for 1993--2003, \cite{cooper} find that the equity risk premium is earned
entirely overnight, and this pattern has continued in the years since. \cite{kellyclark} document
similar differences in 1999--2006 using exchange-traded funds linked to several indices.
\cite{bondarenko} and \cite{boyarch2021} find over-intra outperformance in S\&P 500 index futures.
\cite{bondarenko} explain higher U.S. overnight returns as the result of the resolution of
uncertainty when markets open in Europe.  \cite{boyarch2021} explain it as the result of inventory
management by intermediaries: on days of net selling by investors, dealers absorb greater supply; to
bear the risk of holding this inventory after the market closes, they are compensated with higher
returns.  But the inventory management explanation cannot account for higher returns earned
overnight by stocks that do not lose money during the previous day, nor can it account for
continuation in intraday and overnight returns. We will further contrast our news-based mechanism
with inventory management in Section~\ref{s:reim}.

Inventory management predicts reversals from intraday returns to overnight returns.  \cite{berkman},
and subsequently \cite{aboody}, attribute reversals in the opposite direction to retail investor
sentiment. They find that retail investors bid up the price at the open for the previous day's
high-attention stocks, leading a high overnight return to be followed by a low intraday return.
They measure attention indirectly, through squared returns and net buying. \cite{hrd2024} also
  argue that retail trading is an important contributor to the over-intra effect.  Our news-based
analysis is not necessarily incompatible with an explanation based on retail sentiment; news content
may in fact be a more direct measure of attention to specific stocks. However, our analysis leads to
effective forecasts of {\it annual} overnight and intraday returns which cannot easily be explained
by investor attention over a day or a week.

Several studies contrast anomaly returns intraday and overnight to provide greater insight into the
anomalies.  \cite{lou2019} document the continuation and reversal effects we refer to as over-intra
correlations.  They find that momentum returns are primarily earned overnight, but several other
anomalies (including value, profitability, and investment) earn premia intraday. They relate these
findings to preferences of different types of investors to trade at different times, noting, for
example, that individuals are more likely to trade at the open than institutions (consistent with
\citealt{berkman}), and that institutions trade throughout the day and at the close. These patterns
may indeed be linked, but it is unclear if these timing preferences can explain the over-intra
outperformance and over-intra correlations.  \cite{bogous} notes that capital-constrained
arbitrageurs find it more costly to hold positions (long or short) overnight than intraday, leading
to mispricing at the market close. Like inventory management by dealers, this mechanism may well
contribute to overnight-intraday reversals but it cannot account for the full range of over-intra
outperformance and over-intra correlations.  \cite{akbas2022} find that stocks that experience more
frequent reversals from night to day earn higher returns in the cross section. They conjecture that
reversals are driven by daytime arbitrageurs offsetting the overnight impact of optimistic noise
traders.  Other work relating anomalies to patterns in overnight and intraday returns include
\cite{hendershott} and \cite{lmq}.%
\footnote{In a related paper, \cite{bbrw} propose limiting the calculation of stock liquidity (based
  on \citealt{amihud2002}) to daytime hours because the overnight hours experience significant
  information arrival but little trading.}

Like us, \cite{boudoukh} use textual analysis to study the connection between stock prices and news,
though their methodology is different and they address different questions. They distinguish night
and day as periods driven by public and private information, respectively.  We only consider public
information.  Their main finding, based on overnight returns, is that the arrival of public
information explains a significant fraction of stock return variance. More recently, \cite*{bbm2022}
contend that stock price momentum is driven less by underreaction to public overnight news and more
by underreaction to informed daytime trading. Importantly, they measure momentum returns based on
close-to-close returns.  We find continuation in both intraday and overnight news as measured by
intraday and overnight returns.

\section{Data} \label{s:data}

Our results are based on S\&P 500 firms. Our news data start in 1996 and run through 
December 2022.
Over
this time period, there were 887 firms that were part of the index (based on index constituents from
CRSP) and that were named in news articles within our dataset. We include firms in our analysis only
on days in which they are part of the index.

We obtain daily open and close stock prices from CRSP. We calculate intraday returns from the open
to the close each day; we calculate overnight returns from the close to the next open. Dividends are
paid to the owner of the shares at the close, so we include them in close-to-open returns. We also
assume that all stock splits and buybacks take effect at the close. We calculate an annual intraday
log return as the sum of the individual intraday log returns over a year; annual overnight returns
and monthly intraday and overnight returns are defined similarly.

We calculate various firm characteristics following \cite{green} and using their code, including
size, book-to-market, investment, profitability, and 12-month momentum; we also calculate annualized
volatility.  We apply the inverse hyperbolic sign transformation used in \cite*{glm} to
book-to-market rather than a logarithmic transformation because some book-to-market values are
negative. We calculate 12-month momentum separately from monthly intraday and monthly overnight
returns, in each case leaving out the most recent month and compounding the eleven prior months. We
calculate daily volatilities separately for intraday and overnight returns and annualize them by
multiplying by $\sqrt{252}$.

Our news text comes from the Thomson Reuters News Feed Direct archive (hereafter TR).  Reuters is a
major provider of business news. We follow the pre-processing steps discussed in \cite{glm} for
time-stamping articles, handling chains of articles, and identifying which S\&P 500 companies are
mentioned in an article. We drop articles that do not mention any such companies; we also drop
articles that mention more than three S\&P 500 companies because these usually discuss overall
market developments rather than company-specific news. We drop articles that have fewer than 25
words because these are usually communications included in the TR archive that are not really news
articles.  This process leaves us with roughly 2.4 million news articles, of which approximately
two-thirds are overnight news, in the sense that they are time-stamped after the New York Stock
Exchanges closes at 4 PM and before the next open at 9:30 AM. Articles that fall on weekends and
holidays count as overnight articles for the next trading day. Each intraday and overnight period is
identified by its ending day, i.e., the intraday return is from 9:30 AM to 4 PM on day $s$ but the
day $s$ overnight return is from 4 PM of day $s\!-\!1$ to 9:30 AM on day $s$. The same convention
applies to classifying news articles as intraday or overnight.
 
We apply word stemming to all articles; for example, ``house,'' ``housing,'' and ``housed'' are all
replaced with the token ``hous.''  We remove stopwords, all non-alphanumeric characters, and convert all numbers
to the strings \texttt{\_\_NUM\_\_}, \texttt{\_\_MIL\_\_}, or \texttt{\_\_BIL\_\_}, based on their
magnitude. We focus on single words, and do not analyze bigrams or trigrams.  For our topic
analysis, we drop tokens that appear less than 25 times and in fewer than 25 documents, which leaves us with an effective vocabulary of approximately 44,000 tokens.

\section{Topic model}\label{s:tm}

We use topic modeling to quantify the content of news. Our topic model measures the exposure of
every company to every topic on every day.  These topic exposures are the main tool for our
analysis.  The topics are inferred from the news text (and, in our approach, from stock price
reactions to news); the topics are not imposed exogenously.  We apply a method we introduced in
\cite*{gklm}, which is based on the \emph{latent Dirichlet allocation} (LDA) method of \cite*{lda}.

A topic is defined by a probability distribution over
the effective vocabulary, called the topic-word distribution.  The probability of a word reflects the weight
or importance of that word in the topic.  For example, one topic might assign relatively high
probabilities to ``retail,'' ``sale,'' and ``store,'' and another to ``lawsuit,'' ``claim,'' and
``action.''  Each document is characterized by a probability distribution over the set of topics,
called the document-topic distribution.  This distribution reflects the weight or importance of each
topic for that document.  The topic-word and document-topic distributions are inferred from a
collection of documents.  The user specifies the number of topics $K$ but not the topics
themselves. We take $K\!=\!200$, which worked well in \cite{gklm}; we have also experimented with
100-topic and 500-topic models and found that our results are not very sensitive to this choice.

For purposes of inference, LDA adopts a hypothetical model of how text is generated. The model
posits that each word in a document is selected by first drawing one of the 200 topics from the
distribution over topics specific to that document; then, one of the 44,000 words in the vocabulary
is drawn from the distribution over the vocabulary specific to that topic.  LDA imposes Dirichlet
priors over all of these distributions. Through this specification, the LDA generative model
determines the probability of the words in a collection of documents, given the topic-word and
document-topic distributions. In practice, we observe the words but not the distributions. LDA uses
Bayesian inference to estimate the distributions from the observed words. In short, the model
specifies the probability of text, given the model's distributions; the inference procedure
estimates the distributions, given the observed text.

We use the Gibbs sampling method of \cite{steyvers}, as implemented in \cite{rlda}, to carry out
this inference. The Gibbs sampler repeatedly cycles through our nearly two million news articles, at
each step updating the topic assignment of each word and updating the topic-word and document-topic
distributions. Running through sufficiently many iterations to achieve reasonable convergence of the
topic assignments to their steady state distribution takes several days on a powerful computing
cluster. Section A1 of the Online Appendix contains more details about the topic
model.

In the end, the document-topic distribution for each news article tells us the weight of each topic
in each news article. By summing over all articles that mention a company in a fixed period (a day
or night, for example), we get the weight of each topic in the news about the company.  To make
these weights (or exposures) more explicit, for each article $a$, let
$$
\mbox{Prob}_a(k) = \mbox{probability of topic $k$ in article $a$, for $k=1,\dots,K$;}
$$
these are the document-topic probabilities estimated through LDA.%
\footnote{Here and throughout, the superscript ``$i/o$'' means that there are two separate versions
  of the variable in question, one for the intraday period $i$ and one for the overnight period
  $o$. We also use a superscript $p\in\{i,o\}$ to distinguish between these two periods.}
Let
\begin{equation}
A^{i/o}_{j,t} = \{\mbox{intraday/overnight articles from time $t$ that mention company $j$}\};
\label{aset}
\end{equation}
and let
\begin{equation}
z^{i/o}_{j,k,t} = \sum_{a\in A^{i/o}_{j,t}} \mbox{Prob}_a(k).
\label{zdef}
\end{equation}
The time period $t$ in (\ref{aset}) and (\ref{zdef}) is either a single day (an intraday period
followed by an overnight period) or a year.  If $t$ is a day, the $z^i_{j,k,t}$ is company $j$'s
total exposure to news about topic $k$ released between the open and close on that day.  Similarly,
$z^o_{j,k,t}$ is company $j$'s total exposure to overnight news about topic $k$ from the market
close on day $t\!-\!1$ to the market open on day $t$. When $t$ is a year, (\ref{zdef}) yields the sum of
daily exposures (keeping intraday and overnight news separate) over the year. When we omit the topic
subscript $k$, $z^i_{j,t}$ and $z^o_{j,t}$ each represent 200-dimensional vectors of topic
exposures.  These topic exposure vectors are our main tool for analyzing the content of news and the
market reaction to news.

LDA is an unsupervised method --- it does not have ``true'' topic labels to learn from. The topics
it finds are similar to clusters of co-occurring words. But our interest is in topics that help
explain stock returns. In \cite*{gklm} we introduced a method we called \emph{branching LDA} that
incorporates this consideration.  The method starts by launching ten independent Gibbs
samplers. After a moderately large number of iterations, we compare these models based on their
ability to explain stock market returns. We do this by regressing company returns on the exposure
vectors in (\ref{zdef}) separately for each model and then using each set of regression coefficients
to explain returns from a hold-out sample of news articles. We pick the model with the highest
out-of-sample $R^2$ and discard the others. We then launch 10 new copies of the Gibbs sampler, all
initialized to the state of the first-round winner. We repeat this process for several rounds and
then use the winner of the final round as our topic model. This procedure reinforces choices of
topic-word and document-topic distributions that help explain the stock market's response to
news. As explained in \cite{gklm}, it does so while avoiding the disastrous overfitting of the topic
model that can result from methods that incorporate stock return data into the Gibbs sampler itself.

Using this procedure, we build two topic models: a 2020 (or in-sample) model, and a 2010 (or
out-of-sample) model. The 2020 model uses all our news text from 1996 through 2020 to estimate
document-topic and topic-word distributions. The resulting model would not have been available to an
investor prior to 2020. The 2010 model uses only articles published through 2010 to estimate the
distributions. At the end of 2010, we freeze the document-topic distributions for all pre-2011
articles, and we freeze the topic-word distributions. Using these frozen distributions, we run the
Gibbs sampler on articles published in 2011 and later to infer document-topic distributions for
these articles. This procedure would have been available to an investor to analyze each post-2010
article at the time of its publication. Thus, when we use the 2010 model, the period 2011--2022 is
fully out-of-sample with respect to the topic model.%
\footnote{When we initially ran our topic model, our news data ended in 2020. We have continued to
  use this model for comparison as we acquired two additional years of news.}

We will see examples of specific topics when we discuss Figure \ref{f:10-topic-history} 
in Section~\ref{s:evolution} and Table~\ref{t:company-examples} in Section~\ref{s:examples}.

\subsection{Topic persistence} \label{s:persist}

To quantify the degree of persistence in firm-level news exposures, we run the following panel
regression using the yearly topic vectors from \eqref{zdef}:
\begin{equation} \label{eq:topic-persist}
  z^p_{j,k,t+1} = \alpha_j + \alpha_k + \rho \times z^p_{j,k,t} + \eps^p_{j,k,t+1}\text{,}
\end{equation}
where $\alpha_j$ is a firm fixed effect, $\alpha_k$ is a topic fixed effect, and where
$p \in \{i,o,all\}$ with $z^{all}_{j,k,t}$ defined via
\begin{equation} \label{eq:zall}
  z^{all}_{j,k,t} = z^i_{j,k,t} + z^o_{j,k,t}.
\end{equation}
The $z$'s on the left-hand side of \eqref{eq:topic-persist} are the annual versions of \eqref{zdef}
and the $z$'s on the right-hand side sum over the annual $z$'s in years $\{t-3,\dots,t\}$ and then
divide by 4.  The top panel of Table~\ref{t:persist} shows that the $\rho$ coefficients from the
$z^{all}$ version of \eqref{eq:topic-persist} are just above 0.94. The $p=i$ (intraday) version of
\eqref{eq:topic-persist} yields a $\rho$ coefficient of approximately 0.9 and the $p=o$ (overnight)
version yields a $\rho$ coefficient of around 0.79. Including either a firm or topic fixed effect,
or both, does not change $\rho$ in a meaningful way.

We conclude that, at the yearly level, firm-level topic exposures are very persistent. Companies
whose news load heavily on a given topic in one year tend to have news which load on the same topic
in subsequent years. The effect is particularly strong for intraday news, but also quite pronounced
for overnight news.

\section{News topics and stock returns}

\subsection{Rolling portfolio returns}

\begin{figure}
\centering
\includegraphics[width=5.5in,trim={0.2in 1.25in 2.25in 1.2in},clip]{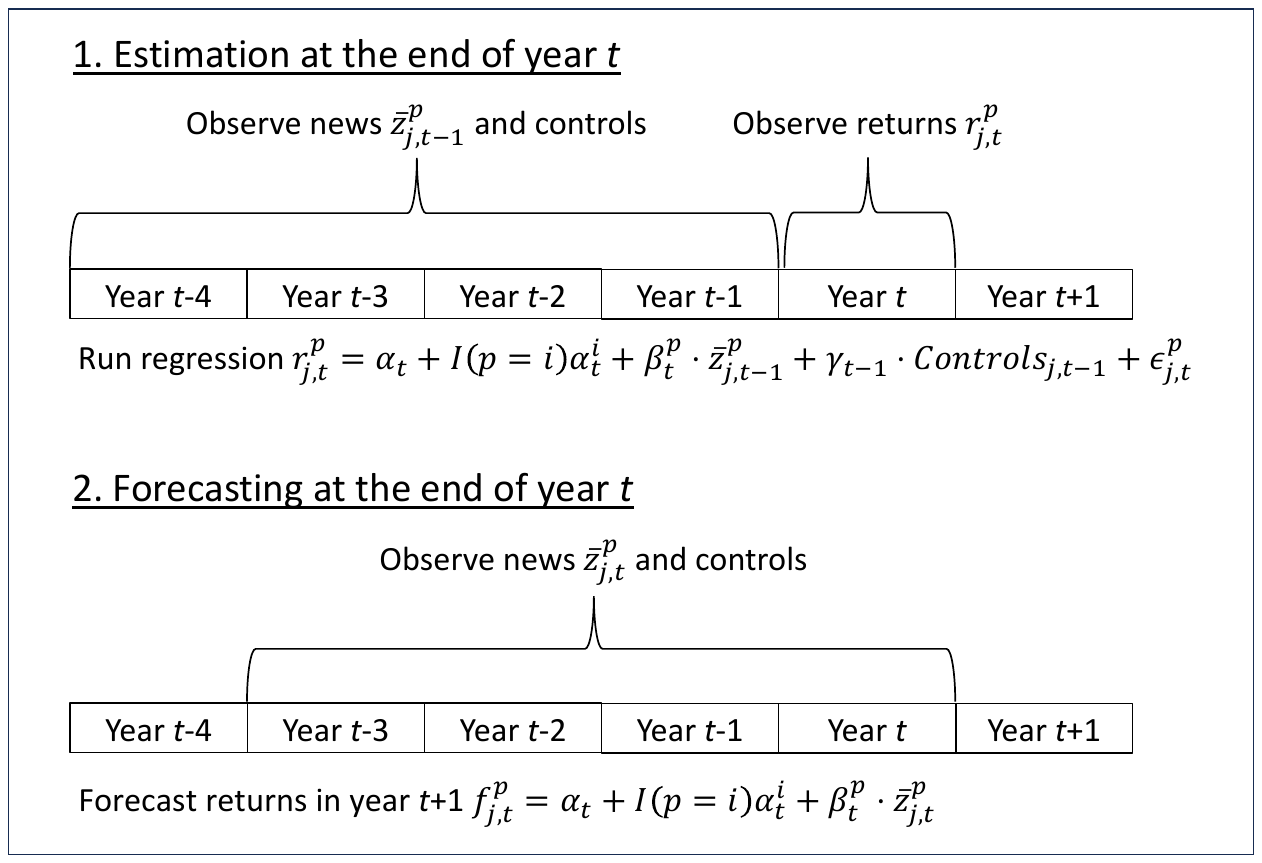}
\caption{Estimation and forecasting using a window of $n=4$ years in (\ref{zsum}) to calculate topic
  exposures $\bar{z}^p_{j,t}$.  At the end of year $t$, we run a (lasso) regression (\ref{rollreg1})
  of returns in year $t$ on topic exposures $\bar{z}^p_{j,t-1}$ and controls from years
  $t-4,\dots,t-1$. In the forecasting step, we apply the coefficients from the estimation step to
  forecast returns in year $t+1$ using topic exposures $\bar{z}^p_{j,t}$, with time index $t$ rather
  than $t-1$, as in (\ref{fdef}). The key point is that we do not use returns or topic exposures
  from year $t+1$ in forming forecasts as of the end of year $t$.}
\label{f:timeline}
\end{figure}

We now investigate the importance of news for over-intra outperformance through rolling
regressions. At the end of each year $t$, we forecast stock returns for the year $t+1$ through the
following steps, which are illustrated in Figure~\ref{f:timeline}.
\begin{itemize}
\item With a window of $n$ years, calculate, for each company $j$,
  \begin{equation}
    \bar{z}^{i/o}_{j,t} = z^{i/o}_{j,t-n+1}  + \cdots + z^{i/o}_{j,t-1} + z^{i/o}_{j,t},
    \label{zsum}
  \end{equation}
  so that $\bar{z}^i_{j,t}$ is the 200-dimensional vector of cumulative intraday topic exposures for
  company $j$ over years $t-n+1$ to $t$, and $\bar{z}^o_{j,t}$ has the corresponding interpretation
  for overnight news.  Our baseline specification uses $n=4$, but we obtain similar results with
  $n=1$.
\item Estimate the coefficient vectors $\beta^i_t$, $\beta^o_t$, and
  $\gamma_t$ in the regressions
  \begin{equation}
    r^p_{j,t} = \alpha_t + I(p=i)\alpha_t^i +
    \beta^p_t\cdot \bar{z}^p_{j,t-1} + \gamma_t\cdot \mathit{CONTROLS}_{j,t-1} + \epsilon^p_{j,t},\quad p\in\{i,o\},
    \label{rollreg1}
  \end{equation}
  where $r^{i/o}_{j,t}$ is the annual intraday or overnight return of company $j$ in year $t$, and
  $I(\cdot)$ denotes the indicator of the event in parentheses, in this case whether a given
  observation is an intraday one.  The controls (discussed in more detail in Section~\ref{s:data})
  are size, book-to-market, investment, profitability, intraday momentum, overnight momentum,
  intraday volatility, and overnight volatility, all measured annually in year $t-1$, and all
  cross-sectionally demeaned within each year.  We use a lasso regression to estimate
  (\ref{rollreg1}) because of the large number of topics. In running the lasso we penalize the
  $\beta^p_t$ vector in \eqref{rollreg1} to retain only the most important topics. We choose the
  penalty parameter through five-fold cross validation to minimize the mean-squared error. The
  $\alpha_t$, $\alpha_t^i$, and $\gamma_t$ are not penalized and are therefore always included in
  the model.  Figure A2 of the Online Appendix shows that the lasso selects
  between 20 and 140 topics per year, with significant variation in this number over time.
  
\item For each company $j$, calculate the intraday and overnight forecasts
  \begin{equation}
    f^p_{j,t} = \alpha_t + I(p=i)\alpha^i_t + \beta^p_t\cdot \bar{z}^p_{j,t},
    \quad p\in\{i,o\}.
    \label{fdef}
  \end{equation}
  Each $f^{i/o}_{j,t}$ forecasts the corresponding return $r^{i/o}_{j,t+1}$ using coefficients
  estimated from past returns and news exposures in years $t-n+1$ to $t$. We omit
  $\mathit{CONTROLS}_{j,t-1}$ from \eqref{fdef} to isolate the predictive power of news flow about
  company $j$ for future intraday and overnight returns.  Note that $\bar{z}^{i/o}_{j,t}$ appears as
  of time $t$ in \eqref{fdef} but in \eqref{rollreg1} it appears as of time $t-1$. Based on our
    topic persistence results in Section \ref{s:persist}, we expect $\bar{z}^p_{j,t}$ to be a good
    proxy for firm $j$'s news flow in year $t+1$. As long as $\beta^p_t$ is also persistent,
    $f^p_{j,t}$ should provide a reasonable forecast for firm $j$'s year $t+1$ stock returns in
    period $p$.
\end{itemize}

As discussed at the end of Section~\ref{s:tm}, we have two versions of our topic model. The 2010
version is estimated using news published through the end of 2010 and then applied to news published
later. Using the 2010 model, the exposure vectors $\bar{z}^{p}_{j,t}$ are available at the end of
year $t$, for $t\ge 2010$, so the forecasts $f^{p}_{j,t}$ in (\ref{fdef}) of the returns
$r^{p}_{j,t+1}$ are fully out-of-sample for $t\ge 2010$. The 2020 model uses articles published
through 2020 to define the topics. Thus, although the forecast (\ref{fdef}) is based on news
articles about company $j$ up to year $t$, the topics that convert these articles into the exposure
vectors $\bar{z}^{p}_{j,t}$ use future news articles. We refer to this case as semi-out-of-sample
because it is out-of-sample with respect to the timing of news flow and stock returns in
(\ref{zsum})--(\ref{fdef}), but it is in-sample with respect to the construction of topics. The
purpose of the semi-out-of-sample case is to see to what extent news can explain past over-intra
differences, regardless of whether the appropriate quantification of news would have been available
to past observers.

We now examine the performance of portfolios formed on the forecasts (\ref{fdef}).  At the end of
each year $t$, we pick the 25 companies $j$ with the highest overnight forecasts $f^o_{j,t}$; call
this the overnight long selected companies LS$^o_t$,
$$
\mbox{LS}^o_t = \{\mbox{companies $j$ with the highest 25 overnight forecasts $f^o_{j,t}$}\}.
$$
(We use LNS$^o_t$ for the other roughly 475 not selected companies.)
We also pick the 25 companies $j$ the lowest intraday forecasts $f^i_{j,t}$; call this is the
intraday short selected companies SS$^i_t$,
$$
\mbox{SS}^i_t = \{\mbox{companies $j$ with the lowest 25 intraday forecasts $f^i_{j,t}$}\}.
$$
(We use SNS$^i_t$ for the other roughly 475 not selected companies.)  We imagine holding the stocks
in LS$^o_t$ only overnight for each day of year $t+1$, earning their equally weighted overnight
return.  Similarly, we hold the stocks in SS$^i_t$ only intraday for each day. With a four-year
window $n=4$ in (\ref{zsum}), trading begins in 2001 because our news archive begins in 1996.%
\footnote{The years 1996-1999 are used to construct $\bar{z}^{i/o}_{j,t}$ in \eqref{zsum} and
  returns from 2000 are used in \eqref{rollreg1}.}

\begin{figure}[b]
\centering
\includegraphics[width=1\linewidth]{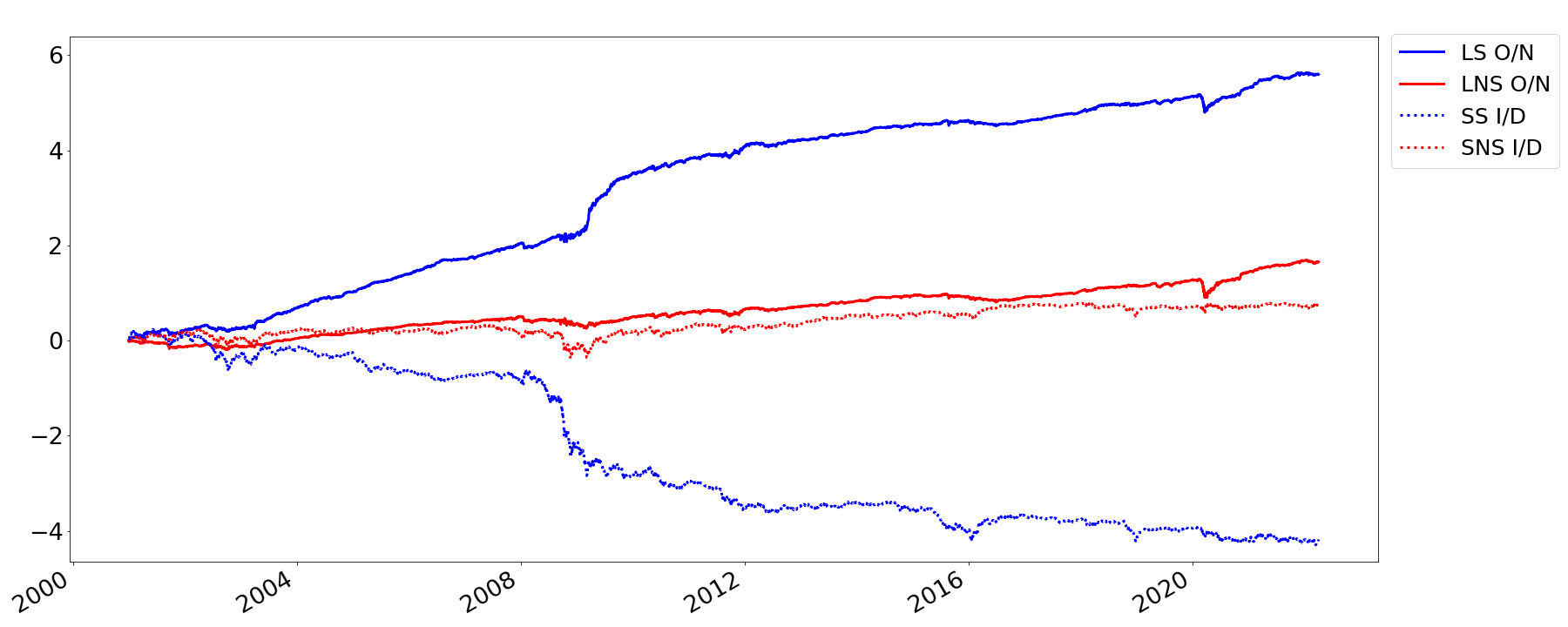}
\caption{The solid blue line shows the cumulative overnight log returns of LS$^o$, the 25 stocks
  with the highest news-based forecast returns. The solid red line shows the cumulative overnight
  log returns of all other stocks. The dotted blue line shows the cumulative intraday log returns of
  SS$^i$, the 25 stocks with the lowest forecast returns. The dotted red line shows the cumulative
  intraday log returns of all other stocks. Stock selections are made annually. Results shown are
  for the 2020 topic model.}
\label{f:varyvary2020}
\end{figure}

The top line of Figure~\ref{f:varyvary2020} shows the cumulative log returns of our overnight long
selected portfolio LS$^o$, based on our 2020 topic model (the semi-out-of-sample case).  The figure
shows phenomenal performance: the stocks predicted to perform well overnight do indeed perform well.
The bottom line of the figure shows the cumulative returns of the intraday short selected portfolio
SS$^i$ and confirms that stocks predicted to perform poorly intraday do indeed perform poorly.%
\footnote{The figure shows the cumulative returns of a long position in SS$^i$. We refer to this
  portfolio as ``short selected'' because its stocks are forecast to do perform poorly, and one
  would want to take a short position in the portfolio to profit from these forecasts.}
In the middle of the figure, the LNS$^o$ line shows the cumulative overnight returns of the 
stocks not selected for LS$^o$, and the SNS$^i$ line similarly shows the cumulative intraday
returns of the stocks not selected for SS$^i$. The non-selected stocks show far less extreme
performance than the selected stocks; moreover, LNS$^o$ and SNS$^i$ perform similarly to each
other. We may summarize these patterns as follows: our news-based forecasts identify the stocks that
perform best overnight and worst intraday; once these stocks are removed, the over-intra
outperformance phenomenon largely disappears.

These comparisons, illustrated graphically in Figure~\ref{f:varyvary2020}, are made precise in the
first row of Table~\ref{t:avg_2020}. The first four columns of the table show the average daily
returns in basis points of the intraday and overnight, selected and non-selected, portfolios.
(Recall that the selections are made at the end of each year and remain fixed throughout the
subsequent year.)  The last four columns test pairwise comparisons.  We calculate standard errors by
regressing returns or differences in returns on a constant using Newey-West standard errors with
auto lag selection.  From these comparisons we see that the selected stocks significantly outperform
the non-selected stocks overnight; they underperfrom intraday; and (in the last column) the
over-intra outperformance is small and not statistically significant once restricted to the
non-selected stocks. In other words, once we remove the stocks forecast to do best overnight and
worst intraday, the over-intra outperformance effect largely disappears.

These results are for the 2020 topic model with returns measured over 2001--2022. The first row of
Table~\ref{t:avg_2011} shows corresponding fully out-of-sample results using the 2010 topic model
with returns measured over 2011--2022.  The results are qualitatively similar to what we saw before;
in particular, the last entry in the row shows that the over-intra outperformance loses significance
once restricted to stocks not selected through their news exposure.%
\footnote{The negative performance of SS$^i$ loses significance in the 2011-2022 period,
but the differences SS$^i$-SNS$^i$ and LS$^o$-SS$^i$ remain highly significant.}
For comparison, the second row of Table~\ref{t:avg_2011} shows corresponding results when we apply
the 2020 topic model with returns measured over 2011--2022. The results are
nearly identical to those for the 2010 model.  Whatever subtle influence post-2010 news may have on
the selection of topics in the 2020 model does not appear to create a performance advantage.  The
semi-out-of-sample and fully out-of-sample results are nearly identical when applied to the same
2011--2022 period.

The last two rows of Table~\ref{t:avg_2011} repeat the analysis of the first two rows using a window
of $n=1$ year for the topic exposure vectors in (\ref{zsum}).  The results are very similar. 

\subsection{Evolution of topics} \label{s:evolution}

Tables~\ref{t:avg_2020}~and~\ref{t:avg_2011} show that time variation in news flow and in market
responses to news flow are important determinants of over-intra outperformance. To gain further
insight into these dynamics, we analyze which topics are the most important determinants of firm
selection into the LS$^o_t$ and SS$^i_t$ sets in a given year. Each firm's year-ahead overnight
return forecast in \eqref{fdef} contains the dot product $\beta^o_t \cdot \bar{z}^o_{j,t}$. For
LS$^o_t$, we consider the Hadamard product (element-by-element multiplication) of the return
sensitivities $\beta^o_t$ with the deviations of topic loadings of the overnight long selected firms
from the universe of all firms:
\begin{equation} \label{eq:contrib-o}
  \phi^o_t \equiv \beta^o_t \odot
  \Big( \underbrace{
    \sum_{j \in \text{LS}^o_t} \bar{z}^o_{j,t} - \bar{z}^o_{SP500,t}
  }_{\text{relative exposures}} \Big),
\end{equation}
where $\bar{z}^o_{SP500,t}$ is the average of $\bar{z}^o_{j,t}$ across all S\&P 500 firms in year
$t$; the units of \eqref{eq:contrib-o} are basis points per year.  This produces a
200-dimensional vector of topic contributions --- each a combination of return sensitivities and
relative topic exposures. For the overnight long selected set, LS$^o_t$, we identify the 10 topics
with the largest positive values in $\phi^o_t$. For the intraday short selected firms, we calculate
$\phi^i_t$ using \eqref{eq:contrib-o} applied to the SS$^i_t$ set, and take the 10 most negative
topics.  For overnight returns, our focus on relative exposures allows a topic to matter even if its
return loading in $\beta^o_t$ is negative, as long as the topic is less prevalent among the selected
firms in LS$^o_t$. Similarly, for intraday returns, topics can be selected even though their
associated $\beta^i_t$ coefficient is positive.

Figure~\ref{f:10-topic-history} shows the 10 most negative topics in $\phi^i_t$ and $\phi^o_t$ for
each year; these use the same topic exposures $\bar{z}^i_{j,t}$ and $\bar{z}^o_{j,t}$ as are used to
form the $t+1$ return forecasts $f^p_{j,t}$ in (\ref{fdef}).%
\footnote{The 25-topic version of Figure \ref{f:10-topic-history}, which is more cluttered but also
  reveals more patterns, is in Figure A3 in the Online Appendix.}
Each contributing topic is summarized with a one- or two-word description, e.g., ``private equity''
refers to the topic whose highest probability stemmed words are: privat, firm, equiti, capit,
partner, fund, invest, buyout, investor, financ. These short descriptions are further grouped into
metatopics. For example, private equity belongs to the metatopic ``Finance \& Investment.''  Our
methodology for summarizing LDA topics and grouping them into metatopics utilizes a large language
model, as described in Section A2 of the Online Appendix.%
\footnote{The time-variation in the importance of topics and metatopics is reminiscent of
  \cite{shiller}'s notion of economic narratives. For more on LDA topics as narratives, see
  \cite{bybee} and \cite{larsen}.}
In some years, fewer than 10 topics are shown because two topics are mapped to the same short
description.  The sign of the $\beta^{i/o}_{t,k}$ coefficient associated with each topic $k$ is
indicated via ``+'' or ``--'' in the figure.

\begin{figure}
  \centering
  {\bf \hspace{0.25in} Overnight vs Intraday Coefficients} \\[5pt]
  \includegraphics[width=1\linewidth]{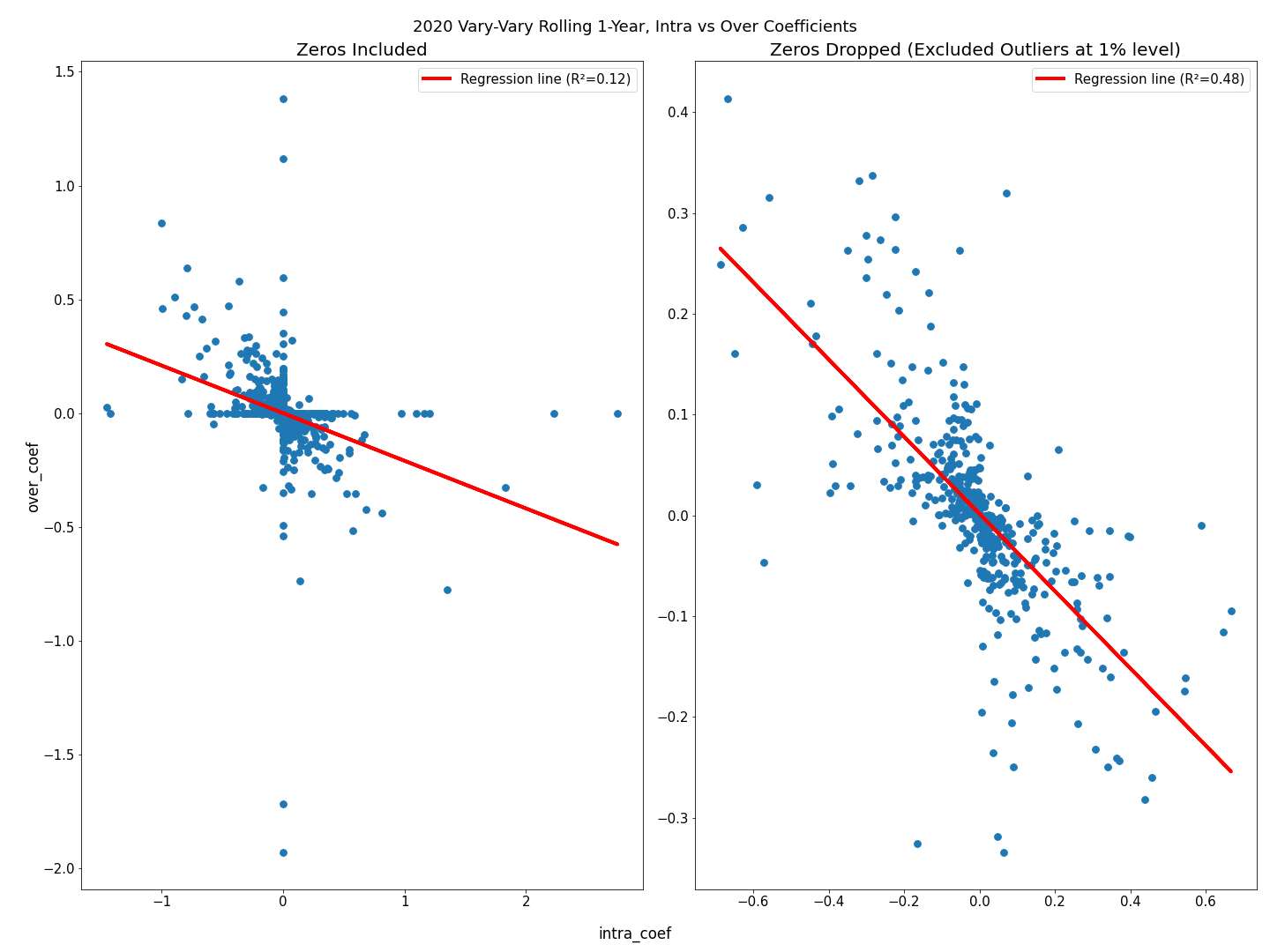}
  \caption{Scatter plots of $(\beta^i_{t,k},\beta^o_{t,k})$, the coefficients in (\ref{rollreg1}),
    for $k$ ranging over topics and $t$ ranging over years 2000 to 2021. The left panel includes all
    coefficients. The right panel shows only the nonzero coefficients (the coefficients for the
    topics selected by the lasso regression when a topic is selected for both intraday and overnight
    returns in a given year and drops extreme values at the 1\% level. Results shown are for the
    2020 topic model.}
  \label{f:over-vs-intra-coef-scatter}
\end{figure}

Panel A of the figure shows the top-contributing intraday topics. For the most part, top intraday
topics are associated with negative $\beta^i_{t,k}$ coefficients in \eqref{fdef} and
\eqref{eq:contrib-o}. Panel B shows the top-contributing overnight topics. Top overnight topics are
mostly associated with positive $\beta^o_{t,k}$ coefficients. The same topic is frequently
identified as important for both intraday and overnight returns in the same year. In such cases,
these topics typically have opposite signs in that year's $\beta^{i/o}_t$ vectors. (Figure
\ref{f:over-vs-intra-coef-scatter} explores this phenomenon further.)

During the global financial crisis, many finance-related topics, e.g., financial, finance, mortgage,
were associated with negative intraday returns and positive overnight returns. In 2003--2005---a
time period that saw several airline bankruptcies---the aviation topic was associated with negative
intraday returns and positive overnight returns. The automotive topic behaved similarly in 2009, the
year in which General Motors and Chrysler filed for bankruptcy. In 2017---the year in which the
Affordable Care Act was enacted by Congress---exposure to the healthcare topic was associated with
negative intraday and overnight returns (which means that firms were selected into LS$^o_{2017}$
based on a lower-than-average exposure to this topic). As a final example, the protection topic
(with top ten most probable words: vaccin, test, covid, coronavirus, health, peopl, virus, infect,
pandem, pfizer) was associated with negative intraday and positive overnight returns in 2020, the
year of the spread of COVID. In summary, a company's loading on topical issues, especially negative
ones, is often associated with negative intraday returns and positive overnight returns in the
year(s) in which the issue is prevalent.

\subsection{Examples of topics and returns} \label{s:examples}

The training and forecasting process in (\ref{zsum})--(\ref{fdef}) suggests the following
interpretation. Suppose a company $j=A$ has a high exposure to some news topic $k$ in years
$t-4,\dots,t-1$ and a large overnight return in year $t$ in (\ref{rollreg1}).  This will contribute
to making the estimated coefficient $\beta_{t,k}^o$ in (\ref{rollreg1}) large and positive. Now
suppose another company $j=B$ has a high exposure to the same topic in years $t-3,\dots,t$. Then the
large coefficient $\beta_{t,k}^o$ and a large topic exposure $\bar{z}^o_{j,t,k}$ will tend to
produce a positive forecast for company $j$'s overnight return in year $t+1$ through
(\ref{fdef}). Thus, we have a positive association between the first company's return $r^o_{A,t}$
and the second company's return $r^o_{B,t+1}$, and this association is mediated by shared exposure
to news topics. It does not rely on serial correlation in one company's return; it finds
cross-company correlations through news flow.

The best evidence for this effect is the performance of the LS$^o_t$ and SS$^o_t$ portfolios because
these portfolios are selected through (\ref{zsum})--(\ref{fdef}).  But we can also illustrate the
phenomenon with specific examples, as in Table~\ref{t:company-examples}. Each subpanel shows a
``spillover'' from a company A on the left to a company B on the right. Consider the first example
in the table. The top row shows that the forecast year $t+1$ is 2008. The topic label is
``collaboration,'' and the metatopic is Corporate Structure \& Business Operations. The next row
shows the highest probability words in the collaboration topic.  These are followed by examples of
news texts that load on the collaboration topic: stories of joint ventures at General Motors and
Nucor. GM's year-$t$ overnight return of $r^o_{A,t}=23.13\%$ is followed by Nucor's year-$(t+1)$
overnight return of $r^o_{B,t+1}=67.77\%$; this is consistent with a positive overnight response to
collaboration news spreading from GM in year $t$ to Nucor in year $t+1$.  Nucor's year-$t$ overnight
return is negative ($r^o_{B,t}=-13.85\%$), ruling out overnight momentum as an explanation for its
year-$(t+1)$ return.

The other examples in Table~\ref{t:company-examples} work similarly. The three subpanels on the left
are based on overnight news and returns, and the three subpanels on the right use intraday news and
returns. In each example, we see an apparent news-mediated positive correlation between $r^p_{A,t}$
and $r^p_{B,t+1}$ that is not shared by $r^p_{B,t}$, which argues against a momentum
effect. Moreover, in all the examples, the A and B companies have little in common beyond their
shared exposure to a news topic. In each case, the direct impact of a single story or topic loading
may be minor, but the cumulative impact of these spillovers produces the large positive returns of
LS$^o_t$ and the large negative returns of SS$^o_t$.

\section{Channels for over-intra outperformance}

The topic summaries of Section \ref{s:evolution} and the examples of Section \ref{s:examples}
demonstrate an important relationship between company news flow and average intraday and overnight
returns. Certain topics in certain years are associated with strongly negative intraday returns, and
some topics are associated with strongly positive overnight returns. The association between such
topic loadings and intraday and overnight returns appears to be very persistent, which explains a
good deal of the over-intra effect. In this section, we evaluate several of the underlying
channels. First, we check whether news variation across intraday and overnight periods, or the
market's response to such news, or both, drives the over-intra effect. Then we analyze the extent to
which influences other than news flow can account for over-intra outperformance.

\subsection{Different news exposures or different market responses?} \label{s:news-or-resp}

We investigate whether the effectiveness of our news-based forecasts is driven by differences in the
content of news intraday and overnight ($\bar{z}^i_{j,t}$ versus $\bar{z}^o_{j,t}$) or by
differences in the response to news intraday and overnight ($\beta^i_t$ versus $\beta^o_t$).  We
will see that variation in the overall exposure to news topics (whether intraday or overnight) and
differences in responses between the two periods are both important in explaining over-intra
outperformance.

To hold the content of news and the response to news fixed across the two time periods, we replace
(\ref{zsum})--(\ref{fdef}) with
\begin{eqnarray}
\bar{z}_{j,t} &=&  \bar{z}^{i}_{j,t} + \bar{z}^{o}_{j,t}
\label{zsum2} \\
r^{p}_{j,t} &=& \alpha_t + I(p=i)\alpha^i_t +\beta_t\cdot \bar{z}_{j,t-1} + \gamma_t\cdot \mathit{CONTROLS}_{j,t-1} + \epsilon^{p}_{j,t}, \quad p\in\{i,o\} 
\label{rollreg2} \\
f^{p}_{j,t} &=& \alpha_t + I(p=i)\alpha^i_t +\beta_t\cdot \bar{z}_{j,t}, \quad p\in\{i,o\}.
\label{fdef2}
\end{eqnarray}
Equation (\ref{zsum2}) combines the intraday and overnight topic exposures, and equation
(\ref{rollreg2}) uses a common $\beta_t$ across the two time periods.  As we did with
(\ref{rollreg1}), we apply lasso regression to (\ref{rollreg2}) (and to equations \ref{rollreg3} and
\ref{rollreg4}, below).

To hold the content of news fixed across the two time periods while allowing different responses, we
replace (\ref{zsum})--(\ref{fdef}) with (\ref{zsum2}) and 
\begin{eqnarray}
r^{p}_{j,t} &=& \alpha_t + I(p=i)\alpha^i_t + \beta^{p}_t\cdot \bar{z}_{j,t-1} + \gamma_t\cdot \mathit{CONTROLS}_{j,t-1} + \epsilon^{p}_{j,t}, \quad p\in\{i,o\} 
\label{rollreg3} \\
f^{p}_{j,t} &=& \alpha_t + I(p=i)\alpha^i_t + \beta^{p}_t\cdot \bar{z}_{j,t}, \quad p\in\{i,o\}.
\label{fdef3}
\end{eqnarray}
Equations (\ref{rollreg3}) and (\ref{fdef3}) allow different responses (as measured by the
coefficients $\beta^{i/o}_t$) to the common exposures in $\bar{z}_{j,t}$.

To allow different topic exposures intraday and overnight while enforcing a common response to these
exposures, we use (\ref{zsum}) as before and replace (\ref{rollreg1}) and (\ref{fdef}) with
\begin{eqnarray}
r^{p}_{j,t} &=& \alpha_t + I(p=i)\alpha^i_t +\beta_t\cdot \bar{z}^{p}_{j,t-1} + \gamma_t\cdot \mathit{CONTROLS}_{j,t-1} + \epsilon^{p}_{j,t}, \quad p\in\{i,o\} 
\label{rollreg4} \\
f^{p}_{j,t} &=& \alpha_t  + I(p=i)\alpha^i_t +\beta_t\cdot \bar{z}^{p}_{j,t}, \quad p\in\{i,o\}.
\label{fdef4}
\end{eqnarray}

Using any of these forecasts, we can form overnight and intraday portfolios and compare their
performance, just as we did in the first row of Table~\ref{t:avg_2020}. The results for
(\ref{fdef2}), (\ref{fdef3}), and (\ref{fdef4}) appear in the next three rows of
Table~\ref{t:avg_2020}, all based on the 2020 model.  In every case, the difference in returns
LS$^o$-SS$^i$ between our overnight and intraday selected stocks is large and statistically
significant, and the last column shows that among the non-selected stocks, the over-intra difference
LNS$^o$-SNS$^i$ loses significance.

Comparing the first two columns across the four cases in Table~\ref{t:avg_2020} indicates that our
overnight selections have higher returns and our intraday selections have more negative returns when
we allow different responses to overnight and intraday news: compare the first and third rows with
the second and fourth. But the results for Model (\ref{rollreg2})--(\ref{fdef2}) show that, even
when we force the intraday coefficients $\beta_t^i$ and overnight coefficients $\beta_t^o$ to
coincide, news exposure helps predict the worst intraday and best overnight performing stocks.%
\footnote{Note that SS$^i$-SNS$^i$ remains significant in the second and fourth rows even
  though SS$^i$ does not.}
Once we force the intraday and overnight coefficients to coincide, separating the intraday and
overnight topic exposures is not important: the results for Model (\ref{rollreg4})--(\ref{fdef4})
are similar to those for Model (\ref{rollreg2})--(\ref{fdef2}). We conclude that the over-intra
outperformance effect is partly explained by different responses to news in the two time periods and
partly explained by a variation in a company's exposure to news topics, regardless of the time of
day of that exposure.

\subsubsection*{Over-intra topic and coefficient correlations}

These patterns are reflected in two correlations: intraday and overnight topic exposures are
positively correlated, whereas intraday and overnight responses to topics are negatively correlated.
Figure~\ref{f:over-vs-intra-coef-scatter} shows scatter plots of coefficient pairs
$(\beta^i_{t,k},\beta^o_{t,k})$, with $k$ ranging over the 200 topics in the 2020 topic model, and
the year index $t$ ranging from 2000 through 2021. The left panel includes all 200 pairs for each
year; topics not selected by the lasso regression have coefficients of zero, which is why many
points in the figure fall on the axes. The right panel includes only the topics selected by the
  lasso for both intraday and overnight returns in a given year; it drops outliers at the top and
bottom 1\% of observations for legibility. In both cases, but especially in the right panel, we see
negative correlation between the intraday and overnight responses to the same topics.

\begin{figure}
  \centering
  {\bf \hspace{0.25in} Overnight vs Intraday Topic Exposures} \\[5pt]
  \includegraphics[width=1\linewidth]{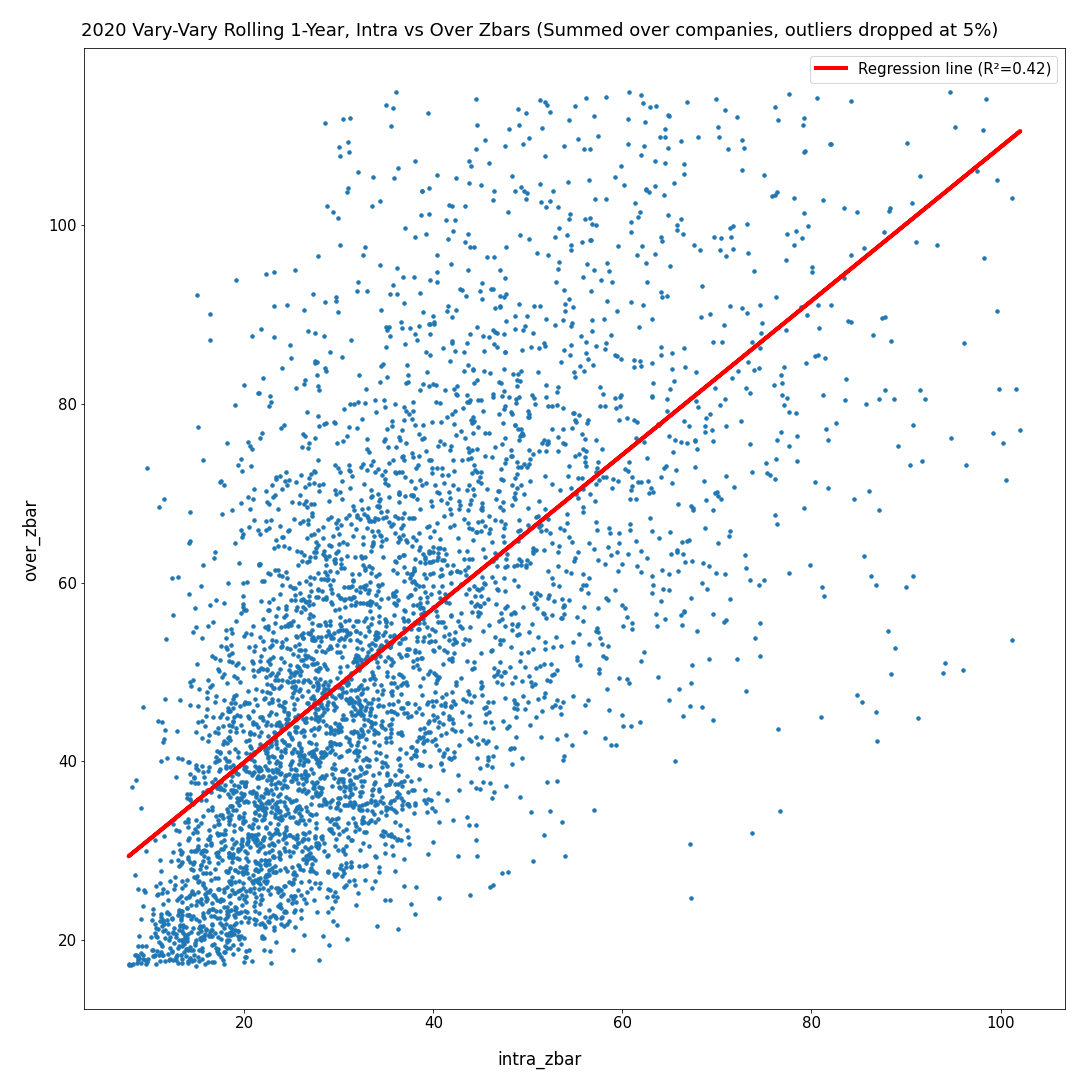}
  \caption{Scatterplot of $(\sum_j {z}_{j,k,t}^i,\sum_j {z}_{j,k,t}^o)$,
where $z^p_{j,k,t}$ is company $j$'s exposure to topic $k$ in year $t$ during period $p\in\{i,o\}$.
The figure drops extreme values at the 5\% level. Results shown are for the 2020 topic model.}
  \label{f:over-vs-intra-zbar-scatter}
\end{figure}

Figure \ref{f:over-vs-intra-zbar-scatter} shows a scatterplot of
$(\sum_j {z}_{j,k,t}^i,\sum_j {z}_{j,k,t}^o)$, where $z^p_{j,k,t}$ is company $j$'s exposure to
topic $k$ in year $t$ during period $p\in\{i,o\}$.  The figure drops outliers at the 5\% level for
legibility. It shows that total exposures to topics (summed over companies) intraday and overnight
are positively correlated.

\begin{figure}
  \centering
  {\bf \hspace{0.25in} Overnight vs Intraday Topic Exposure Correlations} \\[5pt]
  \includegraphics[width=1\linewidth]{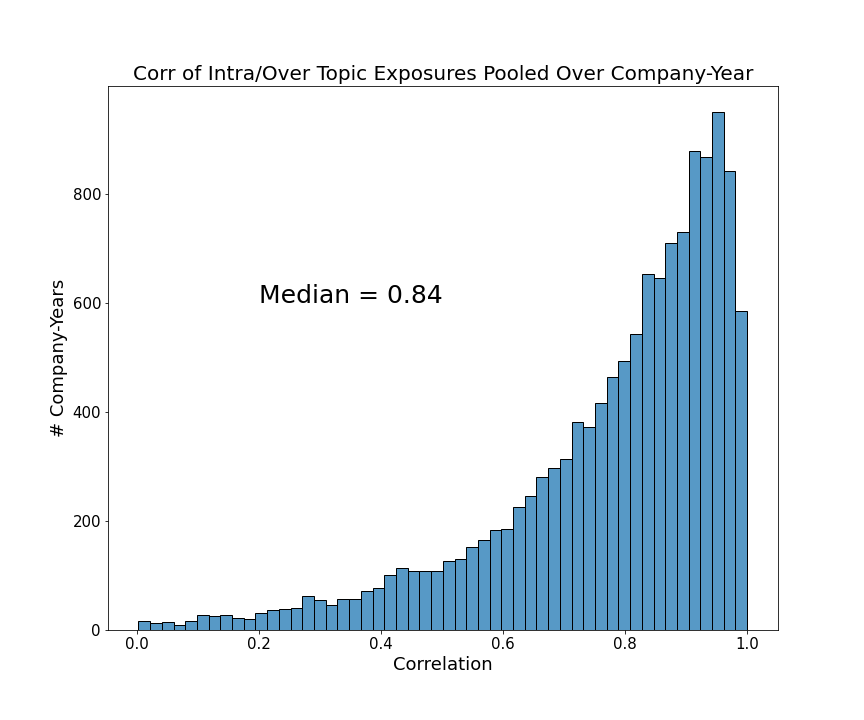}
  \caption{For each (company, year) pair $(j,t)$ we calculate the correlation of exposures $z_{j,t}^i$ and $z_{j,t}^o$ across topics using 1-year exposures. The figure shows the distribution of the correlations obtained in this way.}
  \label{f:over-vs-intra-zbar-scatter-hist}
\end{figure}

In Figure \ref{f:over-vs-intra-zbar-scatter-hist}, we plot the distribution of the correlation of
the pairs $\{(z_{j,k,t}^i,z_{j,k,t}^o),k=1,\dots,200\}$, as $j$ ranges over companies and $t$ over
years.  Whereas Figure \ref{f:over-vs-intra-zbar-scatter} aggregates news exposures over companies,
Figure \ref{f:over-vs-intra-zbar-scatter-hist} measures commonality between an individual company's
intraday and overnight news exposures in a single year. The figure shows that these correlations are
always positive and have a median value of 0.84.

These correlations help explain the comparisons between models in Table~\ref{t:avg_2020}:
the content of intraday and overnight news is positively correlated, so distinguishing news exposures
between the two periods is less important than distinguishing the responses to news, which are
negatively correlated.

\subsection{Characteristics-Adjusted Returns}\label{s:adret}

Tables~\ref{t:avg_2020} and~\ref{t:avg_2011} report average returns. To account for the impact of
firm heterogeneity on overnight versus intraday returns, we now consider returns adjusted for firm
characteristics. We estimate regressions of the form
\begin{eqnarray}
r^p_{j,s} 
&=& b_0 + b_{SNS}\cdot I(p=i,j\in \textrm{SNS}^i_{t(s)-1})
+ b_{LNS}\cdot I(p=o,j\in \textrm{LNS}^o_{t(s)-1}) \nonumber \\
&& +b_{SS}\cdot I(p=i,j\in \textrm{SS}^i_{t(s)-1})
+ c^p\cdot \mathit{CONTROLS}_{j,t(s)-1} + \epsilon^p_{j,s}.
\label{olseval}
\end{eqnarray}
Here,  $r^i_{j,s}$ and $r^o_{j,s}$ are the intraday
and overnight returns, respectively, on day $s$; $t(s)$ is the year containing day $s$; and the
controls are the same demeaned characteristics used in (\ref{rollreg1}).  We also run a baseline
regression 
\begin{equation}
r^p_{j,s} 
=d_0 + d_{intra}\cdot I(p=i)
+ d^p\cdot \mathit{CONTROLS}_{j,t(s)-1} + \tilde\epsilon^p_{j,s}.
\label{baseline}
\end{equation}

Table~\ref{t:adjusted_2020} reports the results for returns in 2001--2022, based on the 2020
topic model. The baseline column reports a value of $-3.6$ bps for $d_{intra}$ in (\ref{baseline});
this is the difference in characteristics-adjusted intraday and overnight returns. Columns (a)--(d)
report estimates of $b_{SNS}$, $b_{LNS}$, $b_{SS}$, and $b_0$ based on
(\ref{olseval}).%
\footnote{Because the controls in (\ref{olseval}) are demeaned, the intercept $b_0$ can
  be interpreted as the mean adjusted return for the selected overnight stock, LS$^o_{t(s)-1}$.}
Each column corresponds to a different way of forming news-based forecasts and thus different
choices of selected sets. Column (d) evaluates selections based on (\ref{rollreg1})--(\ref{fdef}),
which use different intraday and overnight topic vectors $\bar{z}^i$, $\bar{z}^o$ and different
coefficient vectors $\beta^i$, $\beta^o$.  Column (a) evaluates selections based on
(\ref{rollreg2})--(\ref{fdef2}), column (b) corresponds to (\ref{rollreg3})--(\ref{fdef3}), and columns
(c) to (\ref{rollreg4})--(\ref{fdef4}).

In all cases, we find significant outperformance by the overnight selected stocks, as evidenced by
the magnitudes and statistical significance of the intercepts.  The negative and significant
estimates of $b_{LNS}$ indicate that the overnight not-selected stocks underperform the overnight
selected stocks on characteristics-adjusted returns.  The estimates of $b_{SS}$ show that the
intraday selected stocks have significantly negative returns, as predicted.  A comparison of columns
(a) and (c) with columns (b) and (d) suggests that differentiating the response to news intraday and
overnight ($\beta^i$ versus $\beta^o$) has a greater impact than differentiating the content of news
intraday and overnight ($\bar{z}^i$ and $\bar{z}^o$), as we saw previously.

The table also reports p-values for F-tests of equality of
coefficients. The hypothesis that $b_{SS}=b_{SNS}$ is overwhelmingly rejected, showing again that
our intraday selected stocks underperform, as predicted. In contrast, the difference
$b_{SNS}=b_{LNS}$ is largely insignificant: the over-intra outperformance phenonmenon is mostly
eliminated once we remove our news-based picks. This extends our previous finding for average
returns to characteristics-adjusted returns.

Table~\ref{t:adjusted_2011} reports similar results using returns from 2011--2022 rather than
2001--2022.  The first column shows the baseline comparison (\ref{baseline}); the magnitude is
greater than before, but the difference is no longer significant because of the shorter time period
and some increased return volatility.  Column (a) is the same as column (d) of
Table~\ref{t:adjusted_2020} but estimated in the shorter time window; the results are very
similar. Column (b) uses the 2010 topic model and is therefore fully out-of-sample; again, the
results are very similar to what we get for the semi-out-of-sample results in column (a).  Columns
(c) and (d) repeat (a) and (b) with a one-year window for the topic exposures, meaning that we have
$n=1$ rather than $n=4$ in (\ref{zsum}). The results are similar.

\subsection{Removing momentum effects}\label{s:remo}

As discussed in Section~\ref{s:intro}, annual overnight returns and annual intraday returns each
exhibit positive serial correlation.  We have already controlled for these effects by including
overnight and intraday momentum in the controls used in (\ref{rollreg1}) and (\ref{olseval}).
Including momentum in the controls and then omitting the controls when we forecast returns in
\eqref{fdef} orthogonalizes the forecasts with respect to the momentum measures. We now examine two
ways to take this process of removing momentum effects further.

Our first approach randomly splits the set of stocks in half at the end of each year $t$, using a
different randomization each year.  We estimate the regression (\ref{rollreg1}) on one half of the
stocks. We then use those coefficients in (\ref{fdef}) to forecast returns on the other half of the
stocks. We form the LS$^o_t$ and SS$^i_t$ portfolios from these forecasts and evaluate their
performance as before. Separating the stocks used to estimate the market's response to news from the
stocks used to trade on this response removes the potential influence of past returns on our
news-based forecasts. It may also reduce the accuracy of our coefficient estimates (now based on
half as many stocks each year), so even in the absence of momentum effects, we should not expect
performance under this splitting process to match performance based on the full set of stocks.

Column (e) of Table~\ref{t:adjusted_2011} shows the results.  The intraday selected stocks
underperform by even more than before, and the overnight selected stocks (as measured by the
intercept) outperform by even more than before. These results provide strong evidence that
our news-based forecasts are capturing information that is distinct from the serial
correlation in the intraday and overnight returns.

As an alternative approach to further removing momentum effects, we can include the one-day lagged
(and cross-sectionally demeaned) intraday and overnight returns on the
right side of (\ref{olseval}). For intraday returns on day $s$ ($p=i$) we include $r^i_{j,s-1}$ and
$r^o_{j,s}$ as controls, and for overnight returns ($p=o$) we include $r^i_{j,s-1}$ and
$r^o_{j,s-1}$.   Column (f) of Table~\ref{t:adjusted_2011} shows that we get
similar results in this case as well.

\subsection{Removing inventory management effects}\label{s:reim}

\cite{boyarch2021} attribute over-intra outperformance in stock index futures, which they call
overnight drift, to inventory management by intermediaries.  In their account, which draws on the
framework of \cite{grossman}, intermediaries absorb order imbalances at the close of the NYSE, and
they are compensated for this liquidity provision through higher overnight returns. In the specific
case of S\&P 500 E-mini futures studied in \cite{boyarch2021}, much of this overnight return is
earned when European markets open, allowing dealers to offload risk.%
\footnote{Higher overnight returns in U.S. stocks are sometimes attributed to retail investors
  overpaying at the open or overreacting to overnight news. This explanation becomes less plausible
  when the opening price is set by European traders in S\&P 500 futures, as these are less likely to
  be retail investors.}

\cite{boyarch2021} provide compelling evidence for the importance of this mechanism. Even so, this
mechanism cannot account for all aspects of over-intra outperformance and over-intra
correlations. Inventory management ties over-intra outperformance to one-day reversals from intraday
to overnight returns. It does not explain continuation of returns separately in the intraday and
overnight periods, nor does it explain reversals from overnight to intraday returns.  Most
importantly for our investigation, it cannot account for over-intra outperformance that is unrelated
to intra-over reversals, and it is silent on the poor performance of intraday returns.

In this section, we show that the inventory management effect cannot account for the performance of
our news-based picks.  Before presenting new results, we first note that the 50-50 split used in
Section~\ref{s:remo} to remove the influence of return continuation also removes the influence of
cross-period reversals: by separating the stocks used to estimate the responses to news topics from
the stocks used to form news-based portfolios, we ensure that we are not mechanically taking long
overnight positions in stocks that performed poorly intraday in the portfolio formation period.

As a further test, we separate stocks selected by the inventory management effect
from our news-based picks. For each day $s$, let
\begin{equation}
\textrm{IM}_s = \mbox{\{the $m$ stocks with the lowest intraday returns on day $s$\};}
\label{imm}
\end{equation}
we will use $m=25$, 50, and 75. Under the inventory management effect, these are the stocks expected
to have the highest overnight returns $r^o_{j,s+1}$.%
\footnote{This is the close-to-open return for the period ending with the opening of the market on
  day $s+1$.}

We run the following regression of daily overnight returns:
\begin{eqnarray}
  r^o_{j,s+1} 
  &=& b_{\textit Both}\cdot I(j\in \textrm{LS}^o_{t(s)-1})I(j\in \textrm{IM}_{s}) 
      +
      b_{\textit LS-IM}\cdot I(j\in \textrm{LS}^o_{t(s)-1})I(j\not\in \textrm{IM}_{s})
      \nonumber\\
  &&+
     b_{\textit IM-LS}\cdot I(j\not\in \textrm{LS}^o_{t(s)-1})I(j\in \textrm{IM}_{s})
     +
     b_{\textit Rem}\cdot I(j\not\in \textrm{LS}^o_{t(s)-1})I(j\not\in \textrm{IM}_{s})
     \nonumber \\
  && +
     c\cdot \mathit{CONTROLS}_{j,t(s)-1} + \epsilon_{j,s}.
     \label{imeval}
\end{eqnarray}
As before, the controls are demeaned in each year, so the coefficients on the indicators measure
adjusted returns for different portfolios: $b_{Both}$ gives the adjusted return for the stocks
selected both by our news picks and by the inventory managament effect; $b_{LS-IM}$ applies to
stocks selected by news but not by inventory management; $b_{IM-LS}$ applies to stocks selected by
inventory management but not by news; and $b_{Rem}$ applies to the remaining stocks.

Table~\ref{t:IM} shows estimates of $b_{Both}$, $b_{LS-IM}$, $b_{IM-LS}$, and $b_{Rem}$.  The
columns correspond to different choices of $m$ in (\ref{imm}), which is the number stocks selected
by the inventory management strategy. As in previous sections, the LS$^o$ strategy annually picks
the 25 stocks with the highest overnight forecasts.  For $m=25$ and $m=50$, both $b_{LS-IM}$ and
$b_{IM-LS}$ are statistically and economically significant: neither effect is eliminated by removal
of the stocks selected by the other strategy.  At $m=75$ (which leaves few companies
$j\in \textrm{LS}^o_{t(s)-1}$ and $j\not\in \textrm{IM}_{s}$), $b_{LS-IM}$ loses statistical
significance, although its magnitude is only slightly smaller than at $m=50$. Overall, the estimates
of $b_{IS-LM}$ and $b_{IM-LS}$ support the conclusion that our news-based forecasts are using
information that goes beyond the intra-over reversal effect and vice versa.

Interestingly the first row of Table~\ref{t:IM} shows that combining the two signals yields returns
that are greater than the sum of the returns from the isolated signals. This finding reinforces the
conclusion that the two signals carry distinct information.

We also consider an inverse strategy. Whereas the inventory management strategy predicts that stocks
that do poorly intraday will do well overnight, the inverse strategy predicts that stocks that do
well overnight will do poorly intraday. This reversal would be consistent with the finding of
\cite{berkman} that retail investors bid up the opening prices of stocks that receive high attention
the previous day. The higher opening price results in a higher overnight return followed by a lower
intraday return as the overpricing at the open is reversed during the day.

For the inverse strategy define, for each day $s$, 
\begin{equation}
\textrm{IIM}_s = \mbox{\{the $m$ stocks with the highest overnight returns ending on day $s$\};}
\label{iimm}
\end{equation}
we will use $m=25$, 50, and 75. Under the inverse strategy, these are the stocks expected to have
the lowest intraday returns $r^i_{j,s}$.  We replace (\ref{imeval}) with
\begin{eqnarray}
r^i_{j,s} 
&=& b_{\textit Both}\cdot I(j\in \textrm{SS}^i_{t(s)-1})I(j\in \textrm{IIM}_{s}) 
+
b_{\textit SS-IIM}\cdot I(j\in \textrm{SS}^i_{t(s)-1})I(j\not\in \textrm{IIM}_{s})
\nonumber\\
&&+
b_{\textit IIM-SS}\cdot I(j\not\in \textrm{SS}^i_{t(s)-1})I(j\in \textrm{IIM}_{s})
+
b_{\textit Rem}\cdot I(j\not\in \textrm{SS}^i_{t(s)-1})I(j\not\in \textrm{IIM}_{s})
\nonumber \\
&& +
c\cdot \mathit{CONTROLS}_{j,t(s)-1} + \epsilon_{j,s}.
\label{iimeval}
\end{eqnarray}
As in previous sections, the SS$^i$ strategy annually picks the 25 stocks with the lowest forecasts
(\ref{fdef}) of intraday returns.

Table~\ref{t:IIM} shows estimates of $b_{Both}$, $b_{SS-IIM}$, $b_{IIM-LS}$, and $b_{Rem}$.  The
columns correspond to different choices of $m$ in (\ref{iimm}).  In all three cases, both
$b_{SS-IIM}$ and $b_{IIM-SS}$ are statistically and economically significant: neither effect is
eliminated by removal of the stocks selected by the other strategy.  In particular, our news-based
picks underperform intraday (as predicted), even when we remove the previous overnight's best
performers.

The $b_{IIM-SS}$ coefficients have larger magnitudes than the $b_{SS-IIM}$ coefficients, but the
effectiveness of the IIM strategy is not necessarily at odds with our news-based approach. As
already noted, \cite{berkman} associate the reversal from overnight to intraday returns with high
levels of investor attention, with attention measured through volume and volatility. The topic
exposures and topic regression coefficients that drive our return forecasts may themselves be
measures of attention.  We must emphasize, however, that our news-based analysis provides forecasts
of {\it annual} intraday and overnight returns, which cannot be completely explained by investor
attention over a single day.

\subsection{News-based components of intra-over correlations}
\label{s:ioc}

We now examine the influence of news flow on the continuation and reversal effects --- the
intra-over correlations --- introduced in Table \ref{t:io-mom-rev} of Section~\ref{s:intro}.  The
estimates in Table \ref{t:io-mom-rev} show strong continuation effects separately for intraday and
overnight annual returns, and negative contemporaneous correlations of annual intraday and overnight
returns. These effects combine to produce intraday-to-overnight and overnight-to-intraday reversals
in consecutive years. We have considered some of these patterns in our analysis of momentum and
reversal effects in Sections~\ref{s:remo} and~\ref{s:reim}; in this section we examine more directly
the role of news flow in explaining these correlations.

We expect both the content of news and the response to news to contribute to these effects. The
content of intraday and overnight news is persistent (see Table \ref{t:persist}), and the
coefficients (which capture the market's response to news content) tend to be negatively correlated
between the intraday and overnight periods (see Figure \ref{f:over-vs-intra-coef-scatter}).
Taken together, these patterns should help explain positive serial correlation in intraday and overnight
returns and reversals between the two periods.

To measure how news flow contributes to these effects, we run multiple sets of panel regressions. 
For each combination of $p,q\in\{i,o\}$, we run the following three regressions:
\begin{eqnarray}
r^p_{j,t+1} &=& a_t + b_r^{pq}\, r^q_{j,t} + \epsilon^{(r)}_{j,t+1}
\label{rr} \\
r^p_{j,t+1} &=& a_t + c_f^{pq} f^q_{j,t} + \epsilon^{(f)}_{j,t+1}
\label{rf} \\
r^p_{j,t+1} &=& a_t +  b^{pq}\, r^q_{j,t} + c^{pq} f^q_{j,t} + \epsilon_{j,t+1}.
\label{rrf}
\end{eqnarray}
As before, $j$ indexes companies, and $t$ indexes years.
With $p=q$, these regressions test for intra-to-intra or over-to-over correlations;
with $p\not=q$, they test for intra-to-over or over-to-intra correlations.

Consider, for example, the case $p=q=i$. As in (\ref{fdef}), $f^i_{j,t}$ denotes the news-based
forecast of the intraday return for company $j$ over year $t+1$, calculated at the end of year $t$.
The coefficient $b_r^i$ measures the strength of the intraday continuation effect, and the
coefficient $c_f^i$ measures the strength of our news model in forecasting intraday returns. The
first two columns of Table \ref{t:io-panels} show that both coefficients are highly significant. The
third column runs (\ref{rrf}) and finds that both coeffecients remain significant when both
$r^i_{j,t}$ and $f^i_{j,t}$ are included in the regression.  Looking across the subpanels of Table
\ref{t:io-panels}, we see that in every case the coefficients $c_f^{pq}$ (the effect of $f^q_{j,t}$
alone) and $c^{pq}$ (the effect of $f^q_{j,t}$ with $r^q_{j,t}$ included in the regression) remain
significant; and in every case the magnitude of $b^{pq}$ is smaller than the magnitude of
$b_r^{pq}$.  We conclude that our news-based return forecasts explain part of the intra-over
correlations documented in Table \ref{t:io-mom-rev}.

\section{Conclusion}

We show that the outperformance of overnight relative to intraday stock returns is attributable, to
a large extent, to the different sensitivity of daily overnight and intraday returns to past
news. Firm-level news exposures are persistent, which allows us to use a stock's past news exposure,
combined with the market's past response to that news exposure, to estimate how that stock will do
in future intraday and overnight periods. We measure news exposure using a novel topic
classification methodology introduced by \cite{gklm}, which selects topics both for their ability to
capture the probabilistic structure of language, as well as for their ability to explain
contemporaneous stock returns.

We document that companies' loadings on negative topical issues, e.g., the mortgage topic during the
global financial crisis, are often associated with negative intraday and positive overnight
returns. Furthermore, we show that intraday and overnight news coverage tends to be positively
correlated, while the market's response to the same news topic tends to have opposite signs in
intraday and overnight trading periods. This latter effect---the different sensitivity of overnight
and intraday returns to the same news topic---goes a long way to explaining over-intra
outperformance. Controlling for firm characteristics, momentum, and inventory management effects
leaves our main findings unchanged. Finally, we find that stocks experience annual intra-to-intra
and over-to-over return momentum, as well as annual intra-over and over-intra return reversals, an
effect which is also partially explained by companies' overnight and intraday news exposures.

Over-intra outperformance is one of the enduring puzzles in financial markets. 
Our news-based analysis provides a new perspective on this phenomenon.

\appendix

\bibliography{refs_intraover.bib}

\clearpage


\topichistory{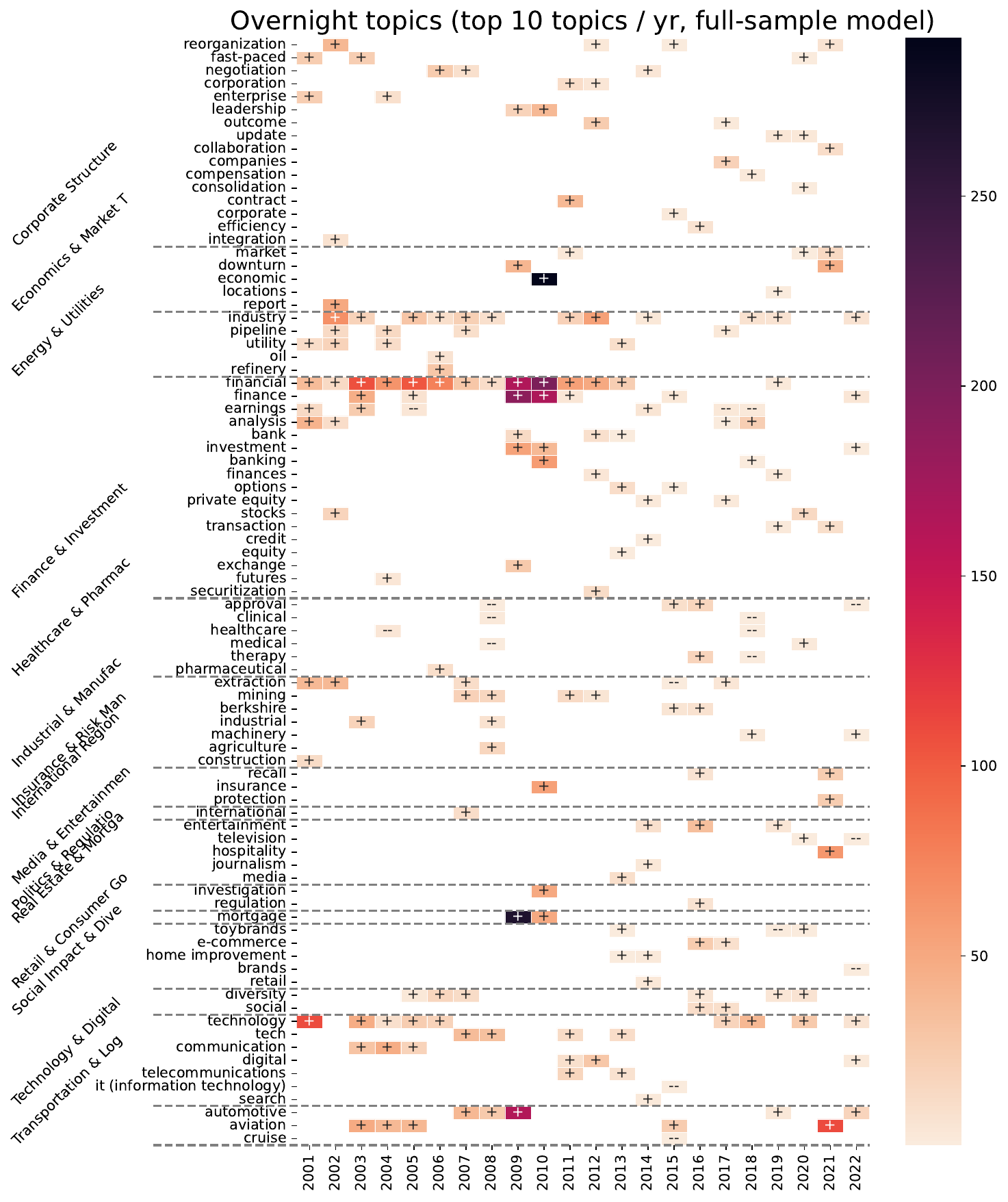}

\clearpage

\begin{table}
  \vspace{-2cm}
  \caption{This table presents variants of the topic persistence specification in
    \eqref{eq:topic-persist}. The first panel uses full-day topic vectors from \eqref{eq:zall}; the
    second panel uses intraday topic vectors; and the third panel uses overnight topic
    vectors. Topic vectors are in annualized terms as explained in Section \ref{s:persist}.
    Standard errors (in parentheses) are double clustered on topic and company for all
    variants.} \label{t:persist} \centering
  {\bf Topic Persistence Regressions} \\
  \begin{tabular*}{\columnwidth}{@{\extracolsep{8pt}}l@{\hskip 0.5in}c@{\hskip 0.5in}c@{\hskip 0.5in}c@{\hskip 0.5in}c} 
    \\[-1.8ex]\cline{1-5}\\[-2.5ex]
    \cline{1-5}
    \\[-1.8ex] & (1) & (2) & (3) & (4) \\[.3ex]
    \cline{1-5} \\
    \underline{All} & 0.947$^{***}$ & 0.944$^{***}$ & 0.943$^{***}$ & 0.943$^{***}$ \\ 
               & (0.025) & (0.026) & (0.027) & (0.027) \\  
               & & & & \\ 
    \cline{1-5} \\
    \underline{Intra} & 0.902$^{***}$ & 0.895$^{***}$ & 0.894$^{***}$ & 0.894$^{***}$ \\ 
               & (0.025) & (0.028) & (0.028) & (0.028) \\ 
               & & & & \\ 
    \cline{1-5} \\
    \underline{Over} & 0.787$^{***}$ & 0.791$^{***}$ & 0.784$^{***}$ & 0.784$^{***}$ \\ 
               & (0.172) & (0.173) & (0.172) & (0.172) \\ 
               & & & & \\ 
    \cline{1-5} \cline{1-5}\\[-1.8ex] 
    Topic fixed effects   & -- & \checkmark & -- &  \checkmark \\ 
    Company fixed effects   & -- & -- & \checkmark  &  \checkmark \\ 
               & & & & \\ 
     \cline{1-5} \cline{1-5}\\[-1.8ex]
    \underline{Adjusted R$^{2}$} & & & & \\
    All & 0.703 & 0.704 & 0.705 & 0.705 \\
    Intra & 0.631 & 0.634 & 0.635 & 0.635 \\
    Over & 0.348 & 0.354 & 0.360 & 0.360 \\
     \cline{1-5} \cline{1-5}\\[-1.8ex]
    Observations &  \multicolumn{4}{c}{1,387,112} \\ 
    \cline{1-5}\\[-2.5ex]
    \cline{1-5}
  \end{tabular*}
\end{table}

\clearpage

\begin{landscape}
{\setlength{\tabcolsep}{10pt}
\begin{longtable}{l@{\hspace{20pt}}lccccccc} 
  \caption{Average daily returns and differences in returns, in basis points, 2001--2022. The
    superscript on each portfolio label indicates whether returns are earned intrday $i$ or
    overnight $o$. LS$^o$ consists of the 25 stocks with the highest forecast overnight returns;
    LNS$^o$ consists of all other stocks. SS$^i$ consists of the 25 stocks with the lowest forecast
    intraday returns; SNS$^i$ consists of all other stocks. Each row uses the indicated forecasting
    model. Forecasts and stock selections are made annually. Portfolio returns are equally
    waited. Results shown are for the 2020 topic model. Average returns are calculated by regressing
    daily returns on a constant. Numbers in parentheses are Newey-West standard errors with
    auto lag selection.} \label{t:avg_2020} \\

  \multicolumn{9}{c}{\bf Average Daily Returns and Difference in Returns (in bps) 2001-2022} \\

  \\[-1.8ex]\hline 
\hline \\[-1.8ex] 
& \multicolumn{8}{c}{Portfolios} \\ 
Forecasts & LS$^o$ & SS$^i$ & LNS$^o$ & SNS$^i$ & LS$^o$ - LNS$^o$ & SS$^i$ - SNS$^i$ & LS$^o$ - SS$^i$ & LNS$^o$ - SNS$^i$ \\
\hline \\[-1.8ex] 
Model  (\ref{rollreg1})--(\ref{fdef})
 & 11.0$^{***}$ & $-$6.3$^{***}$ & 3.4$^{***}$ & 2.0 & 7.6$^{***}$ & $-$8.3$^{***}$ & 17.3$^{***}$ & 1.4 \\  
  & (1.4) & (2.4) & (1.0) & (1.5) & (0.7) & (1.4) & (2.8) & (1.7) \\ 
  & & & & & & & &  \\ 
Model  (\ref{rollreg2})--(\ref{fdef2})
& 5.0$^{***}$ & $-$3.4 & 3.7$^{***}$ & 1.8 & 1.3$^{***}$ & $-$5.2$^{***}$ & 8.4$^{***}$ & 1.9 \\   
  &  (1.1) & (2.4) & (1.0) & (1.5) & (0.4) & (1.5) & (2.7) & (1.7) \\ 
  & & & & & & & &  \\ 
Model  (\ref{rollreg3})--(\ref{fdef3})
& 11.9$^{***}$ & $-$6.4$^{***}$ & 3.3$^{***}$ & 2.0 & 8.6$^{***}$ & $-$8.4$^{***}$ & 18.4$^{***}$ & 1.4 \\  
  & (1.4) & (2.5) & (1.0) & (1.5) & (0.7) & (1.5) & (2.9) & (1.7) \\ 
  & & & & & & & &  \\ 
Model  (\ref{rollreg4})--(\ref{fdef4})
& 5.5$^{***}$ & $-$3.5 & 3.7$^{***}$ & 1.8 & 1.8$^{***}$ & $-$5.3$^{***}$ & 9.0$^{***}$ & 1.9 \\ 
  & (1.2) & (2.4) & (1.0) & (1.5) & (0.5) & (1.5) & (2.7) & (1.7) \\ 
  & & & & & & & &  \\ 
\hline \\[-1.8ex] 
\hline 
\hline \\[-1.8ex] 

\multicolumn{4}{l}{Note: $^{*}$p$<$0.1; $^{**}$p$<$0.05; $^{***}$p$<$0.01} \\ 
\end{longtable}}
\end{landscape}

\clearpage

\begin{landscape}
{\setlength{\tabcolsep}{10pt}
\begin{longtable}{l@{\hspace{20pt}}lccccccc} 
  \caption{Average daily returns and differences in returns, in basis points, 2011--2022. The
    superscript on each portfolio label indicates whether returns are earned intrday $i$ or
    overnight $o$. LS$^o$ consists of the 25 stocks with the highest forecast overnight returns;
    LNS$^o$ consists of all other stocks. SS$^i$ consists of the 25 stocks with the lowest forecast
    intraday returns; SNS$^i$ consists of all other stocks.  Each row uses either the 2010 or 2020
    topic model, with topic exposures calculated over either $n=1$ or $n=4$ years. Forecasts and
    stock selections are made annually based on (\ref{rollreg1})--(\ref{fdef}). Portfolio returns
    are equally waited.  Average returns are calculated by regressing daily returns on a
    constant. Numbers in parentheses are Newey-West standard errors with auto lag selection.}
  \label{t:avg_2011} \\

  \multicolumn{9}{c}{\bf Average Daily Returns and Difference in Returns (in bps) 2011-2022} \\

  \\[-1.8ex]\hline 
\hline \\[-1.8ex] 
 & \multicolumn{8}{c}{Portfolios} \\ 
Topics, $n$ & LS$^o$ & SS$^i$ & LNS$^o$ & SNS$^i$ & LS$^o$ - LNS$^o$ & SS$^i$ - SNS$^i$ & LS$^o$ - SS$^i$ & LNS$^o$ - SNS$^i$ \\
\hline \\[-1.8ex] 
 2010, $n=4$ &  6.9$^{***}$ & $-$3.8 & 4.0$^{***}$ & 2.0 & 2.9$^{***}$ & $-$5.8$^{***}$ & 10.7$^{***}$ & 2.0 \\  
  &  (1.7) & (2.4) & (1.4) & (1.6) & (0.6) & (1.4) & (2.8) & (2.0) \\ 
  & & & & & & & &  \\ 
 2020, $n=4$
 & 6.7$^{***}$ & $-$3.6 & 4.0$^{***}$ & 2.0 & 2.6$^{***}$ & $-$5.6$^{***}$ & 10.3$^{***}$ & 2.1 \\ 
  &  (1.6) & (2.4) & (1.4) & (1.6) & (0.6) & (1.5) & (2.8) & (2.0) \\ 
  & & & & & & & & \\ 
 2010, $n=1$
 & 7.0$^{***}$ & $-$2.5 & 4.0$^{***}$ & 1.9 & 2.9$^{***}$ & $-$4.4$^{***}$ & 9.5$^{***}$ & 2.1 \\ 
  &  (1.6) & (2.2) & (1.4) & (1.6) & (0.6) & (1.1) & (2.7) & (2.0) \\ 
  & & & & & & & & \\ 
  2020, $n=1$ &  7.2$^{***}$ & $-$2.3 & 4.0$^{***}$ & 1.9 & 3.2$^{***}$ & $-$4.2$^{***}$ & 9.5$^{***}$ & 2.1 \\ 
  &  (1.6) & (2.1) & (1.4) & (1.6) & (0.6) & (1.1) & (2.6) & (2.0) \\ 
  & & & & & & & & \\ 
\hline 
\hline \\[-1.8ex] 
 \multicolumn{4}{l}{Note: $^{*}$p$<$0.1; $^{**}$p$<$0.05; $^{***}$p$<$0.01} \\ 
\end{longtable}}
\end{landscape}

\clearpage

\refstepcounter{table}

\begin{twocolumn}[
  \begin{spacing}{1}
    {Table \thetable\label{t:company-examples}:
    Each panel shows the year $t+1$ in the upper-left corner, followed by the short description of
    topic $\tau$, its associated metatopic, followed by the highest probability words in
    $\tau$. The left side of each panel shows the type A company, as well as news text about
    company A from years $t-3,\dots,t$ which loads heavily on topic $\tau$, and company A's year
    $t$ overnight (left column) or intraday (right column) annual return in percent (used
    for model training in equation \ref{rollreg1}). The right side of each panel shows the type B
    company, text from a year $t$ article about company B which loads heavily on topic $\tau$, and
    company B's year $t$ and year $t+1$ (forecast by equation \ref{fdef}) returns in percent.\\[10pt]}
  \end{spacing}
]
\centering

\ABtable
{2008 & collaboration}
{Corp Structure \& Bus Ops}
{ventur, joint, partner, partnership, allianc, agreement, develop, form, statement, strateg, deal}
{GENERAL MOTORS}
{NUCOR}
{A joint venture led by General Motors Corp to revive key assets of South Korea's Daewoo Motor Co will break even in 2005, focusing on exports.} 
{Brazil's CVRD, the world's biggest iron ore producer, on Thursday said it was teaming up with top U.S. steelmaker Nucor Corp. to build a pig iron plant in Northern Brazil.}
{r_{A,t}^o = 23.13}
{r^o_{B,t} = -13.85,\; r^o_{B,t+1} = 67.77}


\ABtable
{2012 & asia}
{International Regions}
{china, chines, hong, kong, singapor, beij, shanghai, yuan, taiwan, firm, asia, net, comreut, countri}
{CITIGROUP}
{CONSOL ENGERGY}
{Shanghai Pudong Development Bank... is based in the commercial hub of Shanghai. Citigroup... has a 4.6\% stake in the lender. The U.S. bank has also agreed to increase its holdings in Pudong Bank to 19.9\%} 
{Consol Energy Inc has sold 88,000 tons of steel-making coal to China in a deal rare for U.S. miners due to shipping distance, a spokesman said on Friday.}
{r_{A,t}^o = 20.70}
{r^o_{B,t} = -59.97,\; r^o_{B,t+1} = 125.21}

\ABtable
{2018 & entertainment}
{Media \& Entertainment}
{disney, cbs, film, walt, netflix, viacom, movi, park, dis, network, studio, abc, star, televis, entertain}
{DISNEY}
{MONDELEZ}
{Disney said on Monday it will produce a new "Star Wars" animated series to air on television in the fall 2014, giving fans of the science-fiction franchise fresh stories before the next live-action film hits theaters in 2015.} 
{Mondelez said it was developing a Sour Patch Kids mobile app, and would change its product advertising by partnering with companies like Twenty-First Century Fox Inc and online publisher Buzzfeed to produce content including video and a live televised event.}
{r_{A,t}^o = 64.89}
{r^o_{B,t} = -8.35,\; r^o_{B,t+1} = 23.04}

\newpage

\ABtable
{2016 & private equity}
{Finance \& Investments}
{privat, firm, equiti, capit, partner, fund, invest, buyout, investor, financ, deal, rais, manag, llc, led}
{APOLLO GROUP}
{APACHE CORP}
{Taminco Group says to be taken over by Apollo Global Management. Taminco, a unit of CVC Capital partners, says to be taken over for about 1.1 billion euros by Apollo.}
{U.S. gas producer Apache is to buy privately owned oil and gas company Cordillera Energy Partners III\dots. Cordillera, owned by private equity firm EnCap Investments
L.P. and other investors, has proven reserves of 71.5 million\dots.} 
{r_{A,t}^i = -5.05}
{r^i_{B,t} = 7.22,\; r^i_{B,t+1} = -21.46}

\vspace*{-9pt}

\ABtable
{2017 & international}
{Politics \& Regulation}
{govern, state, trade, countri, minist, offici, export, foreign, presid, import, world, disput, tariff, against, nation, polit, support, rule, subsidi}
{BOEING}
{FREEPORT MCMORAN}
{The WTO issued confidential findings on Wednesday in one of the world's largest trade disputes, in which the European Union accuses Boeing Co of having benefited from \$5.3 billion subsidies from U.S. authorities.} 
{Indonesia's biggest copper producer has been involved in
lengthy government-led talks aimed at resolving a spat over an
escalating export tax and contract renegotiations, after the
country introduced new export rules}
{r_{A,t}^i = -3.3}
{r^i_{B,t} = 43.67,\; r^i_{B,t+1} = -60.90}
		
\vspace*{-1pt}

\ABtable
{2022 & leadership}
{Corp Structure \& Bus Ops}
{board, ceo, execut, chairman, chief, director, member, retir, resign, appoint, replac, step, presid, serv, offic}
{WELLS FARGO}
{NIKE}
{Wells Fargo \& Co said on Monday Elizabeth Duke has resigned as the chair of its board, effective Mar 8.  James Quigley has also resigned as a member of the board\dots} 
{\dots Nike has experienced nine director-level or higher executive departures over the last 35 days. }
{r_{A,t}^i = -7.91}
{r^i_{B,t} = 21.87,\; r^i_{B,t+1} = -72.82}
\end{twocolumn}

\clearpage

\onecolumn

\begin{table}[!htbp] \centering 
  \caption{Characteristics-adjusted regressions of daily returns 2001--2022, based on the 2020
    model. Numbers in parentheses are standard errors, clustered by time and firm. The $d_{intra}$
    coefficient is from the baseline regression (\ref{baseline}). The coefficients in columns
    (a)--(d) are based on (\ref{olseval}) using portfolio returns from different forecasts: (a) uses
    (\ref{rollreg2})--(\ref{fdef2}), (b) uses (\ref{rollreg3})--(\ref{fdef3}), (c) uses
    (\ref{rollreg4})--(\ref{fdef4}), and (d) uses (\ref{rollreg1})--(\ref{fdef}). The different
    models are also indicated in the row labeled $r^p$. For each model, the table also reports
    p-values for F-tests of equality of the indicated coefficients.}
  \label{t:adjusted_2020}
  {\bf Characteristics-Adjusted Daily Returns Regressions from 2001-2022} \\[5pt]
\begin{tabular}{@{\extracolsep{5pt}}lccccc} 
\\[-1.8ex]\hline 
\hline 
\\[-1.8ex] & baseline & (a) & (b) & (c) & (d)\\ 
\hline \\[-1.8ex] 
$d_{intra}$
  & $-$3.6$^{**}$ &  &  &  &  \\  
  & (1.7) &  &  &  & \\ 
  & & & & & \\ 
$b_{SNS}$
& & $-$5.3$^{***}$ & $-$7.3$^{***}$ & $-$5.4$^{***}$ & $-$7.1$^{***}$ \\  
  & & (2.0) & (2.2) & (1.9) & (2.1) \\ 
  & & & & & \\ 
$b_{LNS}$
& & $-$2.5$^{***}$ & $-$4.7$^{***}$ & $-$2.7$^{***}$ & $-$4.4$^{***}$ \\ 
  & & (0.7) & (0.9) & (0.6) & (0.9) \\ 
  & & & & & \\ 
$b_{SS}$
& & $-$9.2$^{***}$ & $-$14.6$^{***}$ & $-$10.0$^{***}$ & $-$13.7$^{***}$ \\ 
  & &  (2.5) & (2.8) & (2.3) & (2.7) \\ 
  & & & & & \\ 
 $b_0$
 & & 5.0$^{***}$ & 7.1$^{***}$ & 5.2$^{***}$ & 6.9$^{***}$ \\ 
  & & (1.3) & (1.5) & (1.2) & (1.4) \\ 
  & & & & & \\ 
\hline \\[-1.8ex] 
p-values for tests
 & & & & & \\
\cline{1-1} \\[-1.8ex]
$b_{SNS} = b_{LNS}$
& & 0.103 & 0.142 & 0.113 & 0.134 \\  
$b_{SNS} = b_{SS}$
& & 0.000 & 0.000 & 0.000 & 0.000 \\  \\ \hline \\[-1.8ex] 
Type $p\in\{i,o\}$ & & $\beta\cdot \bar{z}$ & $\beta^p\cdot\bar{z}$ & $\beta\cdot\bar{z}^p$ &  $\beta^p\cdot\bar{z}^p$ \\
\\[-1.8ex] \hline \\[-1.8ex]
Observations & \multicolumn{5}{c}{5,036,404} \\ 
\hline \\[-1.8ex] 
\textit{Note:}  & \multicolumn{5}{r}{$^{*}$p$<$0.1; $^{**}$p$<$0.05; $^{***}$p$<$0.01} \\ 
\end{tabular} 
\end{table} 

\clearpage 

\begin{longtable}{@{\extracolsep{0pt}}lccccccc} 
  \caption{Characteristics-adjusted regressions (\ref{olseval}) of daily returns 2011--2022. Numbers
    in parentheses are standard errors, clustered by time and firm. The $d_{intra}$ coefficient is from
    the baseline regression (\ref{baseline}). Columns (a) and (b) evaluate returns based on
    (\ref{rollreg1})--(\ref{fdef}) using the 2020 and 2010 models, respectively. Columns (c) and (d)
    repeat (a) and (b) with news exposures $\bar{z}_{j,t}$ measured over $n=1$ year; all other
    results in the table use $n=4$. In generating returns for column (e), we randomly split the set
    of stocks each year and use half to estimate $\beta^p_t$ in (\ref{rollreg1}) and the other half
    to form portfolios based on (\ref{fdef}). For column (f), we include lagged returns
    $r^i_{j,s-1}$ and $r^o_{j,s-1}$ on the right side of (\ref{olseval}). Columns (e) and (f) use
    the 2010 topic model.}\label{t:adjusted_2011} \\

  \multicolumn{8}{c}{\bf Characteristics-Adjusted Daily Returns Regressions from 2011-2022} \\

  \\[-1.8ex]\hline 
\hline \\[-2.2ex] 
 & & \multicolumn{2}{c}{$n=4$} & \multicolumn{2}{c}{$n=1$} & 50-50 Split & With Lags\\ 
 & & 2020 & 2010 & 2020 & 2010 & 2010 & 2010\\ 
\cline{3-4} \cline{5-6} \cline{7-7} \cline{8-8}
\\[-1.8ex] & baseline & (a) & (b) &   (c) & (d) & (e) & (f)  \\ 
\hline \\[-2.2ex] 
$d_{intra}$
& $-$3.1 & & & & & &\\ 
 & (2.0) & & & & & &\\ 
  & & & & & & &  \\ [-2.2ex]
 $b_{SNS}$
 & & $-$4.6$^{**}$ & $-$4.7$^{**}$ & $-$4.2$^{*}$ & $-$4.3$^{**}$ & $-$5.6$^{**}$ & $-$4.9$^{**}$ \\  
  & & (2.3) & (2.4) & (2.2) & (2.1) & (2.4) & (2.4) \\ 
  & & & & & & &  \\ [-2.2ex]
$b_{LNS}$
& & $-$2.0$^{***}$ & $-$2.2$^{***}$ & $-$1.5$^{**}$ & $-$1.7$^{***}$ & $-$2.9$^{***}$ & $-$2.4$^{***}$ \\ 
  & & (0.7) & (0.8) & (0.6) & (0.6) & (1.0) & (0.8) \\ 
  & & & & & & &  \\ [-2.2ex]
$b_{SS}$
& & $-$8.4$^{***}$ & $-$9.8$^{***}$ & $-$7.7$^{***}$ & $-$7.5$^{***}$ & $-$8.0$^{***}$ & $-$10.1$^{***}$ \\ 
  & & (2.8) & (3.0) & (2.6) & (2.5) & (2.9) & (3.0) \\  
  & & & & & & &  \\ [-2.2ex]
$b_0$
& & 5.2$^{***}$ & 5.4$^{***}$ & 4.8$^{***}$ & 4.9$^{***}$ & 6.1$^{***}$ & 5.6$^{***}$ \\ 
  & & (1.7) & (1.9) & (1.7) & (1.5) & (1.9) & (1.9) \\  \\ 
  & & & & & & &  \\ [-2.2ex]
\hline \\[-2.2ex] 
p-values for tests  & & & & & \\ 
\cline{1-1} \\[-2.2ex]
$b_{SNS}=b_{LNS}$
& & 0.199 & 0.220 & 0.190 & 0.187 & 0.194 & 0.213  \\ 
$b_{SNS} = b_{SS}$
& & 0.002 & 0.000 & 0.001 & 0.002 & 0.098 & 0.000 \\  \hline \\[-2.2ex] 
Type $p\in\{i,o\}$ & \multicolumn{7}{c}{$\beta^p\cdot\bar{z}^p$} \\
\\[-2.2ex] \hline \\[-2.2ex]
Observations & \multicolumn{7}{c}{2,627,660} \\ 
\hline 
\hline \\[-2.2ex] 
\textit{Note:}  & \multicolumn{7}{r}{$^{*}$p$<$0.1; $^{**}$p$<$0.05; $^{***}$p$<$0.01} \\ 
\end{longtable} 

\clearpage

\begin{table}[!htbp] \centering 
  \caption{Removing the inventory management effect. The columns show coefficient estimates from
    (\ref{imeval}) using different values for $m$, which is the number of stocks selected by the
    inventory management (intraday to overnight reversal) strategy. Numbers in parentheses are
    standard errors, clustered by time and firm.}
  \label{t:IM} 
  {\bf Removing the Inventory Management Effect} \\[5pt]
\begin{tabular}{@{\extracolsep{5pt}}lccc} 
\\[-1.8ex]\hline 
\\[-1.8ex] &  $m=25$ & $m=50$ & $m=75$ \\ 
\hline \\[-1.8ex] 
 $b_{Both}$ & 10.0$^{***}$ & 11.5$^{***}$ & 10.7$^{***}$ \\ 
  & (3.3) & (2.5) & (2.3) \\ 
  & & & \\ 
 $b_{LS-IM}$ & 3.8$^{**}$ & 3.1$^{*}$ & 2.7 \\  
 &  (1.7) & (1.7) & (1.7) \\ 
  & & & \\ 
$b_{IM-LS}$
& 5.2$^{**}$ & 6.1$^{***}$ & 6.1$^{***}$ \\ 
  & (2.1) & (1.8) & (1.6) \\ 
  & & & \\ 
  $b_{Rem}$ & 3.2$^{**}$ & 3.0$^{**}$ & 2.8$^{**}$ \\ 
  & (1.4) & (1.4) & (1.4) \\  
  & & &\\ 
\hline \\[-1.8ex] 
Observations & \multicolumn{3}{c}{1,313,830}  \\ 
\hline 
\hline \\[-1.8ex] 
\textit{Note:}  & \multicolumn{3}{r}{$^{*}$p$<$0.1; $^{**}$p$<$0.05; $^{***}$p$<$0.01} \\ 
\end{tabular} 
\end{table} 

\begin{table}[!htbp] \centering 
  \caption{Removing the inverse inventory management effect. The columns show coefficient estimates
    from (\ref{iimeval}) using different values for $m$, which is the number of stocks selected by
    the inverse inventory management (overnight to intraday reversal) strategy. Numbers in
    parentheses are standard errors, clustered by time and firm.}
  {\bf Removing the Inverse Inventory Management Effect} \\[5pt]
  \label{t:IIM}
\begin{tabular}{@{\extracolsep{5pt}}lccc} 
\\[-1.8ex]\hline 
\\[-1.8ex] &  $m=25$ & $m=50$ & $m=75$ \\ 
\hline \\[-1.8ex] 
 $b_{Both}$ &  $-$12.2$^{*}$ & $-$8.9$^{*}$ & $-$6.9$^{*}$ \\ 
  &  (6.3) & (4.5) & (3.9) \\ 
  & & & \\ 
 $b_{SS-IIM}$ & $-$3.8$^{*}$ & $-$3.8$^{*}$ & $-$3.9$^{*}$ \\ 
 & (2.2) & (2.2) & (2.2) \\ 
  & & & \\ 
$b_{IIM-SS}$
&  $-$9.6$^{***}$ & $-$6.2$^{***}$ & $-$4.3$^{**}$ \\ 
  & (2.3) & (2.0) & (1.9) \\ 
  & & & \\ 
  $b_{Rem}$ &  1.1 & 1.4 & 1.5 \\ 
  & (1.6) & (1.6) & (1.6) \\ 
  & & &\\ 
\hline \\[-1.8ex] 
Observations & \multicolumn{3}{c}{1,313,830}  \\ 
\hline 
\hline \\[-1.8ex] 
\textit{Note:}  & \multicolumn{3}{r}{$^{*}$p$<$0.1; $^{**}$p$<$0.05; $^{***}$p$<$0.01} \\ 
\end{tabular} 
\end{table}

\clearpage

\begin{landscape}
\begin{table}
\centering
\caption{News-based components of intra-over correlations. Results use the 2020 topic model with
  annual returns for 2000--2022. Each subpanel of three regressions corresponds to an instance of
  (\ref{rr})--(\ref{rrf}), $p,q\in\{i,o\}$, with time fixed effects. Standard errors (in
  parentheses) are clustered by time and company.} \label{t:io-panels}
{\bf News-Based Components of Intra-Over Correlations} \\[5pt]
\begin{tabular}{r|lll|lll|lll|lll}
\toprule \toprule
& \multicolumn{3}{c|}{Intra to Intra} & \multicolumn{3}{c|}{Over to Intra} & \multicolumn{3}{c|}{Over to Over} & \multicolumn{3}{c}{Intra to Over}\\
{} & $r^i_{t+1}$ & $r^i_{t+1}$ & $r^i_{t+1}$ & $r^i_{t+1}$ & $r^i_{t+1}$ & $r^i_{t+1}$ & $r^o_{t+1}$ & $r^o_{t+1}$ & $r^o_{t+1}$ & $r^o_{t+1}$ & $r^o_{t+1}$ & $r^o_{t+1}$ \\
\midrule
$r^i_t$     &      0.191$^{***}$ &             &      0.155$^{***}$ &             &             &             &             &             &             &     -0.172$^{***}$ &             &     -0.145$^{***}$ \\
 &       (0.042) &             &       (0.035) &             &             &             &             &             &             &       (0.049) &             &     (0.046) \\
$r^o_t$     &             &             &             &     -0.348$^{***}$ &             &     -0.313$^{***}$ &      0.290$^{***}$ &             &      0.257$^{***}$ &             &             &             \\
 &             &             &             &      (0.053) &             &       (0.053) &       (0.044) &             &       (0.035) &             &             &             \\
$f^i_t$     &             &      0.079$^{***}$ &      0.039$^{**}$ &             &      0.079$^{***}$ &      0.042$^{*}$ &             &             &             &             &             &             \\
 &             &       (0.023) &     (0.017) &             &       (0.023) &     (0.022) &             &             &             &             &             &             \\
$f^o_t$     &             &             &             &             &             &             &             &      0.067$^{***}$ &      0.028$^{***}$ &             &      0.067$^{***}$ &      0.045$^{***}$ \\
 &             &             &             &             &             &             &             &       (0.014) &     (0.009) &             &       (0.014) &       (0.009) \\
  \midrule
Obs        &       10084 &       10084 &       10084 &       10084 &       10084 &       10084 &       10084 &       10084 &       10084 &       10084 &       10084 &       10084 \\
\bottomrule \bottomrule
\end{tabular}
\vskip 5pt
\textit{Note:} $^{*}$p$<$0.1; $^{**}$p$<$0.05; $^{***}$p$<$0.01
\end{table}
\end{landscape}

\end{document}

%% file: common_defs.tex
\usepackage{xr}
\usepackage{setspace}
\usepackage{booktabs}
\usepackage{hyperref}
\usepackage{amsmath}
\usepackage{xcolor}
\usepackage{natbib}
\usepackage{amsthm}
\usepackage{float}
\usepackage{graphicx}
\usepackage{subcaption}
\usepackage{pdflscape}
\usepackage{amssymb}
\usepackage{lscape}
\usepackage{longtable}
\usepackage{rotating}
\usepackage{multirow}
\usepackage{tabularx}
\usepackage{pdfpages}
\usepackage{caption}
\usepackage{adjustbox}

\newcommand{\commentall}[1]{}
\newcolumntype{C}{@{\extracolsep{.5cm}}c}%
\newcolumntype{L}[1]{>{\flushleft\arraybackslash}p{#1}}

\def\paperTitle{Does Overnight News Explain Overnight Returns?}
\def\authorLine{\author{Paul Glasserman\thanks{Columbia Business School, pg20@columbia.edu.}
    \and Kriste Krstovski\thanks{Columbia Business School, kriste.krstovski@columbia.edu.}
    \and Paul Laliberte\thanks{Columbia Business School, pl2669@columbia.edu.}
    \and Harry Mamaysky\thanks{Columbia Business School, hm2646@columbia.edu.}}}

\def\eps{\epsilon}

\newcommand\topichistory[1]{
  \begin{landscape}
    \begin{figure}
      \centering
      {\bf \hspace{1.1in} Panel A \hspace{4in} Panel B} \\[3pt]
      \mbox{
        \includegraphics[height=0.85\textheight]
        {images/main_paper/meta_topic_plots/Intraday_topics_top_#1_topics_per_yr_full-sample_model}
        \includegraphics[height=0.85\textheight]
        {images/main_paper/meta_topic_plots/Overnight_topics_top_#1_topics_per_yr_full-sample_model}
      }
      \caption{Evolution of intraday and overnight topic forecast contributions in
        \eqref{eq:contrib-o}. The sign of the $\beta^{i/o}_t$ coefficient associated with each
        selected topic is indicated via ``+'' or ``--''. The color of each rectangle shows the
        magnitude of the elements of the $\phi^{i/o}_t$ vectors, in basis points per year when
        summed over all companies belonging to either LS$^o_t$ or SS$^i_t$ in \eqref{eq:contrib-o}.}
      \label{f:#1-topic-history}
    \end{figure}
  \end{landscape}
}

\newcommand{\ABtable}[9]{%
\begin{table}[h!]
{\scriptsize
\begin{tabular}{|cccc|}\hline
#1 &	\multicolumn{2}{c|}{#2}\\
\hline
\multicolumn{4}{|p{3.in}|}{#3}
\\ \hline
\multicolumn{2}{|c|}{#4} & \multicolumn{2}{|c|}{#5} \\
\multicolumn{2}{|p{1.3in}|}{#6}&
\multicolumn{2}{|p{1.8in}|}{#7}
\\ \hline
\multicolumn{2}{|l|}{$#8$} & \multicolumn{2}{|l|}{$#9$}
\\ \hline
\end{tabular}
}
\end{table}
}